\documentclass[apj]{emulateapj}
\pdfoutput=1

\newcommand{\chandra}{{\it Chandra}}

\newcommand{\rosat}{{\it ROSAT\,}}
\newcommand{\asca}{{\it ASCA}}
\newcommand{\euve}{{\it EUVE}}
\newcommand{\exosat}{{\it EXOSAT}}
\newcommand{\einstein}{{\it Einstein}}

\tightenlines
\slugcomment{To appear in the {\em Astrophysical Journal}}
\lefthead{Drake et al.}
\righthead{\chandra\ HRC Observations of AR Lac}

\begin{document}

\title{A 33 year constancy of the X-ray coronae of AR Lac and eclipse diagnosis of scale height}

\author{Jeremy J. Drake,\altaffilmark{1} Peter Ratzlaff,\altaffilmark{1} Vinay Kashyap,\altaffilmark{1} David P. Huenemoerder,\altaffilmark{2} Bradford J.~Wargelin,\altaffilmark{1} Deron~O.~Pease\altaffilmark{3}}
\affil{$^1$Harvard-Smithsonian Center for Astrophysics, 
\\ 60 Garden Street, Cambridge, MA 02138}
\affil{$^2$Massachusetts Institute of Technology, Kavli Institute for Astrophysics and Space Research,\\ 77 Massachusetts Avenue, Cambridge, MA 02139}
\affil{$^3$University of California, Berkeley,
Space Sciences Laboratory,\\
7 Gauss Way,
Berkeley, CA 94720}

\begin{abstract}
Extensive X-ray and extreme ultraviolet (EUV) photometric observations
of the eclipsing RS~CVn system AR~Lac were obtained over the years
1997 to 2013 with the {\it Chandra X-ray Observatory} {\it Extreme
Ultraviolet Explorer}.  During primary eclipse, HRC count rates
decrease by $\sim$40\%.  A similar minimum is seen during one primary
eclipse observed by \euve\ but not in others owing to intrinsic source
variability.  Little evidence for secondary eclipses is present in
either the X-ray or EUV data, reminiscent of earlier X-ray and EUV
observations.  Primary eclipses allow us to estimate the extent of a
spherically symmetric corona on the primary G star of about $1.3
R_\odot$, or $0.86 R_\star$, and indicate the G star is likely
brighter than the K component by a factor of 2--5.  Brightness changes
not attributable to eclipses appear to be dominated by stochastic
variability and are generally non-repeating.  X-ray and EUV light
curves cannot therefore be reliably used to reconstruct the spatial
distribution of emission assuming only eclipses and rotational
modulation are at work.  Moderate flaring is observed, where count
rates increase by up to a factor of three above quiescence. Combined
with older \asca, \einstein, \exosat, \rosat\ and {\it Beppo-SAX} observations, the
data show that the level of quiescent coronal emission at X-ray wavelengths
has remained remarkably constant over 33 years, with no sign of variation due to 
magnetic cycles. Variations in base level X-ray emission seen by {\it
Chandra} over 13 years are only $\sim 10$\%, while variations back to
pioneering {\it Einstein} observations in 1980 amount to a maximum of
45\%\ and more typically about 15\%.
 \end{abstract}

\keywords{stars: activity --- coronae --- late-type --- individual: AR Lac
binaries: close --- X-rays: stars}

\section{INTRODUCTION}


AR Lac is the brightest known totally eclipsing system of the RS~CVn
class of close binaries.  RS~CVn systems have orbital periods
typically between 1 and 14 days, with orbital separations of only a
few stellar radii \citep{Hall:78}.  Tidal viscosity tends to
synchronize the stellar rotation and orbital periods, increasing the
rotation rates beyond typical values for all but the youngest single
late-type stars.  This rapid rotation is thought to engender strong
dynamo action that makes them magnetically active and thus
copious sources of chromospheric and coronal emission; they are among
the very brightest stellar coronal sources observed at short
wavelengths with X-ray luminosities up to ten thousand times that of
the typical active Sun
\citep[e.g.][]{Walter.etal:78,Pallavicini.etal:81,Drake.etal:92,Dempsey.etal:93,Singh.etal:96,Makarov:03,Pandey.Singh:12}.

Only about 1\% of the Sun's surface is covered by bright active region emission
during solar maximum \citep{Drake.etal:00}, 
so even if the entire solar surface were covered in such emission, 
the RS~CVn systems would still be two orders of magnitude more luminous in X-rays.
These higher luminosities
could be achieved through higher plasma densities, larger radial
extent, or a mixture of both.  
It seems likely that these very active
coronae include a continuously flaring component
\citep[see also][]{Guedel:97,Drake.etal:00}.
As an eclipsing
RS~CVn-type system, AR~Lac has played a key role in attempts to
understand the morphology of these very active stellar coronae.

The first clues to the spatial structure of the AR Lac coronae came
from radio observations: \citet{Owen.Spangler:77} failed to detect an
eclipse in the quiescent radio emission at 4585~MHz, a result which
suggested that the radio flux originates from a region that is large
compared to the radii of the component stars.
%
%
From radio interferometry, \citet{Trigilio.etal:01}
determined that the emission was spatially resolved and of order of
the binary dimensions; slight variability outside of eclipses suggested
modulation by inhomogeneous structures.

Further progress, though with partially conflicting results, was made
through direct observations of the hot coronal plasma by the {\it
  Einstein} \citep{Walter.etal:83}, {\it EXOSAT}
\citep{White.etal:90,Siarkowski:92}, {\it ROSAT} \citep{Ottmann.etal:93}, \asca\
\citep{White.etal:94,Siarkowski.etal:96}, {\it EUVE} \citep{Walter:96,
  Christian.etal:96}, and {\it Beppo-SAX} \citep{Rodono.etal:99}
observatories.
The lack of an obvious eclipse in the harder of two \exosat\ X-ray
bandpasses lead \citet{White.etal:90} to suggest the hotter and cooler
plasma resides in two distinct regions, with the harder emission
coming from a much larger region, comparable to the size of the
stellar system.  This conclusion was bolstered by a similar \exosat\ observation showing an apparently uneclipsed hot component on the active binary TY~Pyx \citep{Culhane.etal:90}, and fitted well with both the
\citet{Owen.Spangler:77} radio result and the finding of
\citet{Swank.etal:81} that low resolution {\it Einstein} spectra of
active stars could be adequately fitted with discrete two-temperature
models containing a hard and softer component.  Subsequent EUV and
X-ray studies
all observed distinct primary eclipses\footnote{We adopt here the usual convention for AR Lac designating the G2~IV star to be the primary.}, supporting the view that
a significant fraction of AR Lac coronal emission must arise
from a relatively compact region.
Detailed reviews and
discussion of these different observations have been presented by
\citet{Christian.etal:96} and \citet{Rodono.etal:99} .

Further insights were made spectroscopically, first with {\it EUVE},
and then with the {\it Chandra} and {\it XMM-Newton} observatories,
both of which are equipped with diffraction gratings
permitting detailed high-resolution spectroscopy in the soft X-ray
bandpass \citep[e.g.][]{Weisskopf.etal:03,den_Herder.etal:01}.
\citet{Griffiths.Jordan:98} estimated a coronal plasma density of
$5\times 10^{11}$~cm$^{-3}$ based on Fe~XXI lines.
\citet{Huenemoerder.etal:03} also found tentative evidence for high
plasma densities on the order of $\log n_e \sim 11$~cm$^{-3}$ using
lines from the He-like ions of O and Ne formed around 2--$4\times
10^6$~K.  This result was confirmed by \citet{Testa.etal:04}, who
found $\log n_e =12.5\pm 0.5$~cm$^{-3}$ from He-like Mg formed at
slightly hotter temperatures of $\sim 6\times 10^6$~K \citep[see also][]{Ness.etal:04}.
\citet{Testa.etal:04} also analysed the {\it Chandra} spectra of
several other active binaries and found similar high densities to
generally characterise the coronae of very active stars.  High
densities point toward compact coronae, though without further spatial
diagnostics there remains a degeneracy between surface filling factor
and coronal scale height.

The degeneracy was broken for two RS~CVn-type binaries stars, II~Peg,
IM~Peg, and the active M~dwarf EV~Lac, that were part of a larger
sample whose high-resolution {\it Chandra} HETG spectra were analysed
by \citet{Testa.etal:04,Testa.etal:07}.  These stars exhibited
significant resonance scattering depletion of H-like O and Ne
Ly$\alpha$ lines.  The size of coronal structures derived from the
measured optical depths for all three sources is of the order of a few
percent of the stellar radius at most, indicating the presence of
compact, dense and very bright emitting structures.

One drawback of existing X-ray studies of AR~Lac is that data
generally cover only fractions of an orbit, or concentrate on one
orbital period or less.  It is difficult to tell from all the
disparate observations what the long-term behaviour of the source is,
how repeatable any eclipses are, and what emission variations are
likely to be due to rotational modulation or to simple intrinsic
stochastic variability.

The high elliptical orbit of the {\it Chandra X-ray Observatory},
combined with high spatial resolution and relatively low noise
detectors, provides an advantageous viewpoint for studying the
time-dependent X-ray emission of stellar coronae.  During on-orbit
calibration toward the end of the summer of 1999, AR~Lac was 
favorably placed in the sky and of the right X-ray brightness to make
a suitable point-source calibration target for the {\it Chandra} High
Resolution Camera (HRC).  As such, it was the first X-ray bright
late-type star observed by {\it Chandra}, and has been observed
regularly since then to monitor instrument performance.  Here
we present an analysis of these data that were 
obtained over a period of 13 years from 1999 to 2012 and represent the most
extensive set of observations of the coronae of AR~Lac yet undertaken.

Following a brief summary of the adopted parameters of AR Lac in
\S\ref{s:arlac}, in \S\ref{s:obs} we describe the observational
material and data reduction; \S\ref{s:anal} and \S\ref{s:discuss} present  
an analysis
and discussion of the light curves and their implication for the
structure of the AR Lac coronae, as well as the coronae of similarly
active stars; a summary and conclusions are presented in
\S\ref{s:sum}.

\section{AR LAC}
\label{s:arlac}

AR~Lac lies at a distance of 42~pc \citep[e.g.][]{Siviero.etal:06},
has a period of 1.983 days and comprises G2 IV and K0 IV stars of
approximately equal masses but unequal radii, 
separated by a distance of
about 9.2~$R_{\odot}$ \citep{Chambliss:76,Popper.Ulrich:77}.  In the optical band, 
the G star is completely eclipsed by the K0
subgiant.  

We adopt the system parameters of \citet{Popper:90} and the ephemeris of
\citet{Siviero.etal:06} that is based on an extensive set of optical eclipse
observations.  For our purposes, this ephemeris is essentially identical to that of \citet{Marino.etal:98}, that was also adopted by \citet{Rodono.etal:99}.  The relevant parameters are 
listed in Table~\ref{t:arlac}.  While \citet{Lu.etal:12} have more recently studied the orbital period variation of AR~Lac, and have produced an analytical formula for the difference between observed and predicted eclipse times (``$O-C$'') relative to the ephemeris of  \citet{Siviero.etal:06}, these $O-C$ corrections do not seem to match the data of Siviero et al., with values of $O-C$ close to 0.1 days for the range of epochs on which Siviero et al.\ base their ephemeris.  We also find that the \euve\ and \chandra\ eclipses reported here are not consistent with the \citet{Lu.etal:12} $O-C$  values.  

\section{OBSERVATIONS AND LIGHT CURVES}
\label{s:obs}

\subsection{Chandra X-ray Photometry}
\label{s:chandra}

Regular {\it Chandra} observations of AR Lac have been performed using
the High Resolution Camera imaging and spectroscopic (HRC-I, HRC-S)
detectors since the initial on-orbit calibration phase in 1999.  These
data were obtained for the purposes of verifying the focus and imaging
performance of the combined mirror and detector assembly.  The HRC
instrument is of microchannel plate design with sensitivity in the 0.07-10~keV range and peaking around 1~keV, 
and provides photon timing
resolution of a few $ms$ in the standard mode employed for the data presented here 
(and up to 16~$\mu s$ in a special ``timing mode'' \citealt{Kenter.etal:00}) that enable accurate light curves to be
constructed.  Since the detectors themselves have only very low energy resolution, no attempt was made to constrain or filter detected events in energy. 

\begin{deluxetable}{lcc}
\tablecaption{Relevant parameters of AR Lac adopted in this study \citep[from][]{Popper:90} \label{t:arlac}}
\tablehead{
\colhead{} &
\colhead{Primary} &
\colhead{Secondary} \\
\colhead{} &
\colhead{G2 IV} &
\colhead{K0 IV} }
\startdata
 Mass   & $1.23 M_\odot$ &$1.27 M_\odot$ \\
  Radius & $1.52 R_\odot$ & $2.72 R_\odot$  \\
Inclination & \multicolumn{2}{c}{$i=87\deg$} \\
Primary eclipse\tablenotemark{a} & \multicolumn{2}{c}{2451745.58650+1.98318608E} \\ 
 \enddata
 \tablenotetext{a}{From \citet{Siviero.etal:06}}
\end{deluxetable}

\begin{deluxetable*}{llllll}
\tablecaption{Chandra HRC Observations of AR~Lac used in this Analysis\label{t:hrcobs}}
\tablehead{
\colhead{Observation} &
\colhead{Detector} &
\colhead{UT First Start} &
\colhead{UT Last Stop} &
\colhead{Elapsed} &
\colhead{Exposure} \\
\colhead{ID} &
\colhead{} &
\colhead{} &
\colhead{} &
\colhead{(ks)} &
\colhead{(ks)}
}

\startdata

1283-1289, 1294-1295 & I & 1999-08-31T19:31:51 & 1999-09-01T00:41:32 & 19 & 7 \\
1319-1382, 1385 & I & 1999-10-03T13:10:26 & 1999-10-05T05:34:48 & 145 & 103 \\
1484-1504 & I & 1999-12-09T09:41:42 & 1999-12-09T20:40:04 & 40 & 25 \\
996, 2345-2364 & I & 2000-12-12T16:31:38 & 2000-12-13T01:37:11 & 33 & 25 \\
998, 2366-2385 & S & 2000-12-20T14:52:41 & 2000-12-20T23:51:27 & 32 & 24 \\
997, 2432-2451 & S & 2001-05-14T00:05:25 & 2001-05-14T10:30:09 & 37 & 28 \\
2625-2645 & S & 2002-01-26T14:09:49 & 2002-01-26T23:03:17 & 32 & 24 \\
2604-2624 & I & 2002-01-26T23:03:17 & 2002-01-27T08:01:37 & 32 & 25 \\
2646-2666 & S & 2002-08-09T11:30:11 & 2002-08-10T12:52:59 & 91 & 18 \\
4332-4352 & S & 2003-02-22T00:26:08 & 2003-02-22T09:26:05 & 32 & 24 \\
4290-4310 & I & 2003-02-22T09:26:05 & 2003-02-22T18:23:36 & 32 & 24 \\
4311-4331 & S & 2003-09-01T09:42:51 & 2003-09-01T18:59:46 & 33 & 23 \\
5081-5101 & S & 2004-02-09T12:59:22 & 2004-02-09T21:39:16 & 31 & 22 \\
5060-5062 & I & 2004-09-13T20:19:58 & 2004-09-13T21:39:48 & 5 & 3 \\
5063-5080, 6133-6135 & I & 2004-11-25T13:40:27 & 2004-11-25T22:21:40 & 31 & 22 \\
5102-5122 & S & 2004-11-28T05:42:35 & 2004-11-28T14:13:21 & 31 & 22 \\
6021-6041 & S & 2005-02-10T10:38:01 & 2005-02-10T20:24:24 & 35 & 26 \\
6000-6020 & S & 2005-09-01T20:58:49 & 2005-09-02T06:40:12 & 35 & 26 \\
5979-5989 & I & 2005-09-27T08:06:24 & 2005-09-27T13:38:26 & 20 & 7 \\
5996-5997 & I & 2005-10-02T19:10:59 & 2005-10-02T20:12:35 & 4 & 2 \\
5990-5992 & I & 2005-10-09T14:54:37 & 2005-10-09T16:47:07 & 7 & 2 \\
5993-5995, 5998-5999 & I & 2005-10-17T18:19:18 & 2005-10-17T23:35:24 & 19 & 6 \\
6477-6497 & S & 2006-03-20T05:05:39 & 2006-03-20T15:02:41 & 36 & 26 \\
6519-6539 & I & 2006-09-20T19:20:57 & 2006-09-21T05:06:40 & 35 & 27 \\
6498-6518 & S & 2006-09-21T18:56:18 & 2006-09-22T04:51:09 & 36 & 26 \\
8298-8318 & I & 2007-09-17T13:08:38 & 2007-09-17T22:40:41 & 34 & 27 \\
8320-8340 & S & 2007-09-21T17:06:23 & 2007-09-22T03:05:09 & 36 & 26 \\
9682-9683 & S & 2008-07-11T07:47:29 & 2008-07-11T09:55:09 & 8 & 6 \\
9684-9685 & I & 2008-07-11T09:55:09 & 2008-07-11T11:53:54 & 7 & 6 \\
9661-9681 & S & 2008-09-02T02:37:56 & 2008-09-02T13:31:27 & 39 & 30 \\
9640-9660 & I & 2008-09-07T09:35:46 & 2008-09-07T20:03:22 & 38 & 30 \\
10578-10598 & I & 2009-09-24T16:07:52 & 2009-09-25T01:53:30 & 35 & 26 \\
10601-10621 & S & 2009-09-25T21:51:14 & 2009-09-26T07:35:29 & 35 & 27 \\
11889-11909 & I & 2010-09-25T02:40:08 & 2010-09-25T12:09:40 & 34 & 27 \\
11910-11930 & S & 2010-09-25T12:09:40 & 2010-09-25T21:42:13 & 34 & 26 \\
13182 & I & 2010-12-16T18:45:33 & 2010-12-17T00:14:08 & 20 & 18 \\
13265, 13048-13067 & I & 2011-09-18T20:48:16 & 2011-09-19T06:06:52 & 34 & 26 \\
13068-13088 & S & 2011-09-19T06:06:52 & 2011-09-19T15:49:50 & 35 & 26 \\
14278-14298 & S & 2012-09-24T09:42:17 & 2012-09-24T19:23:23 & 35 & 26 \\
14299-14319 & I & 2012-09-27T02:28:47 & 2012-09-27T12:14:25 & 35 & 27 \\
15409-15429 & I & 2013-09-16T15:20:29 & 2013-09-18T06:39:54 & 142 & 27 \\
15430-15450 & S & 2013-09-16T18:11:51 & 2013-09-18T08:31:45 & 138 & 26 \\

\enddata
\end{deluxetable*}

The AR~Lac observations were aimed at different off-axis angles to
obtain pointings over a range of detector locations in a coarse
``raster'' , with each pointing typically lasting from one to a few ks
each.  Each pointing has associated ``start'' and ``stop'' times
separated by short intervals during which the detector high voltage
was ramped down.  The observations are summarised in
Table~\ref{t:hrcobs}, though details of the individual pointings within each visit are omitted.  In a small handful of cases, individual pointings were found to have dithered onto the High Energy Suppression Filter attached to the HRC-S and were discarded.

Satellite telemetry was processed by standard {\it Chandra X-ray
  Center} (CXC) pipeline procedures to produce photon event lists.
Raw instrument count rates were examined to ensure that the data were
not affected by telemetry saturation, which can lead to significant
deadtime.  Times of telemetry saturation and where the instrument dead time might be significant (as judged by the flag DTF$ < 0.98$) were discarded.  


\begin{figure*}
\begin{center}
\includegraphics[angle=270,width=5.5in]{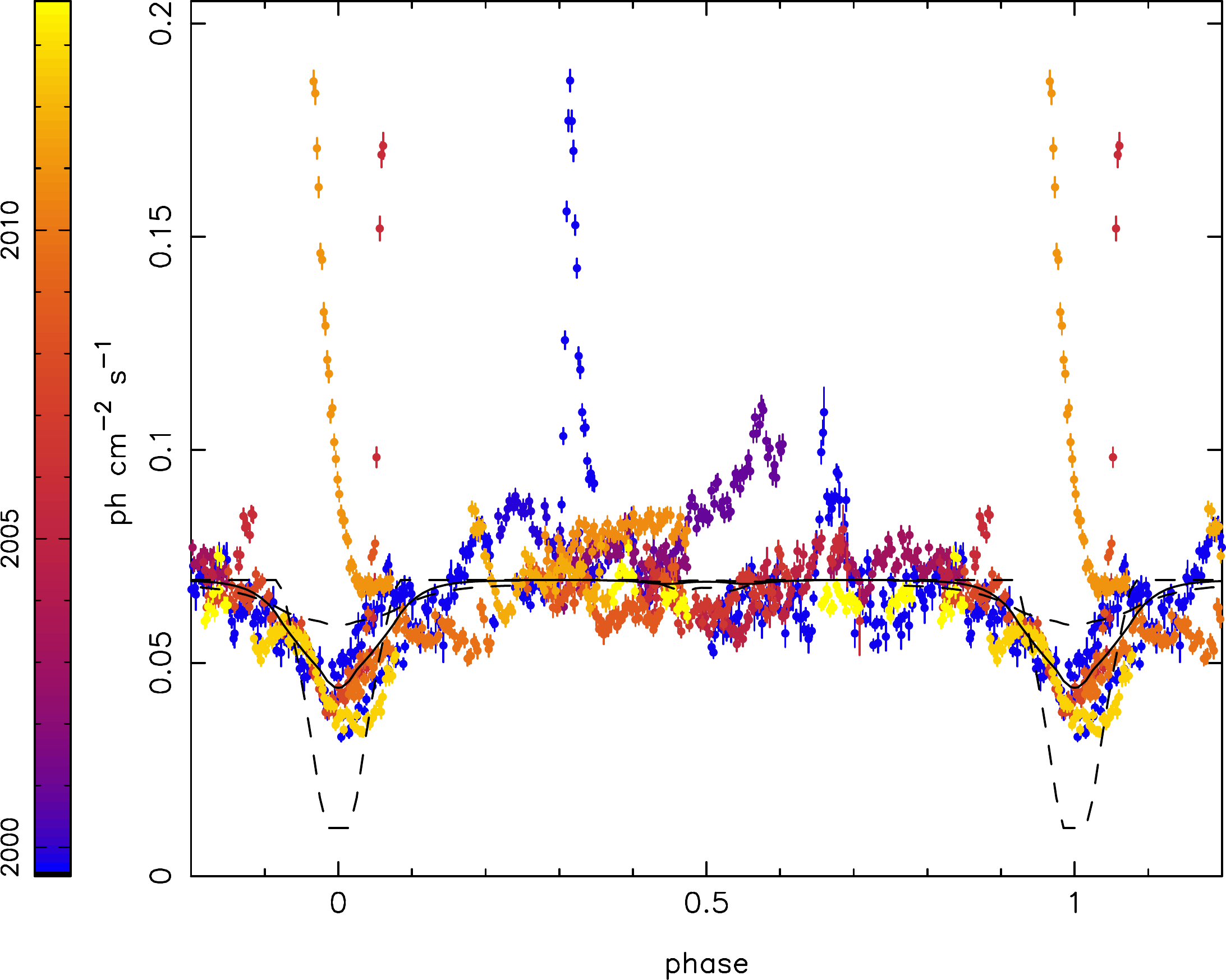}
\includegraphics[angle=270,width=5.5in]{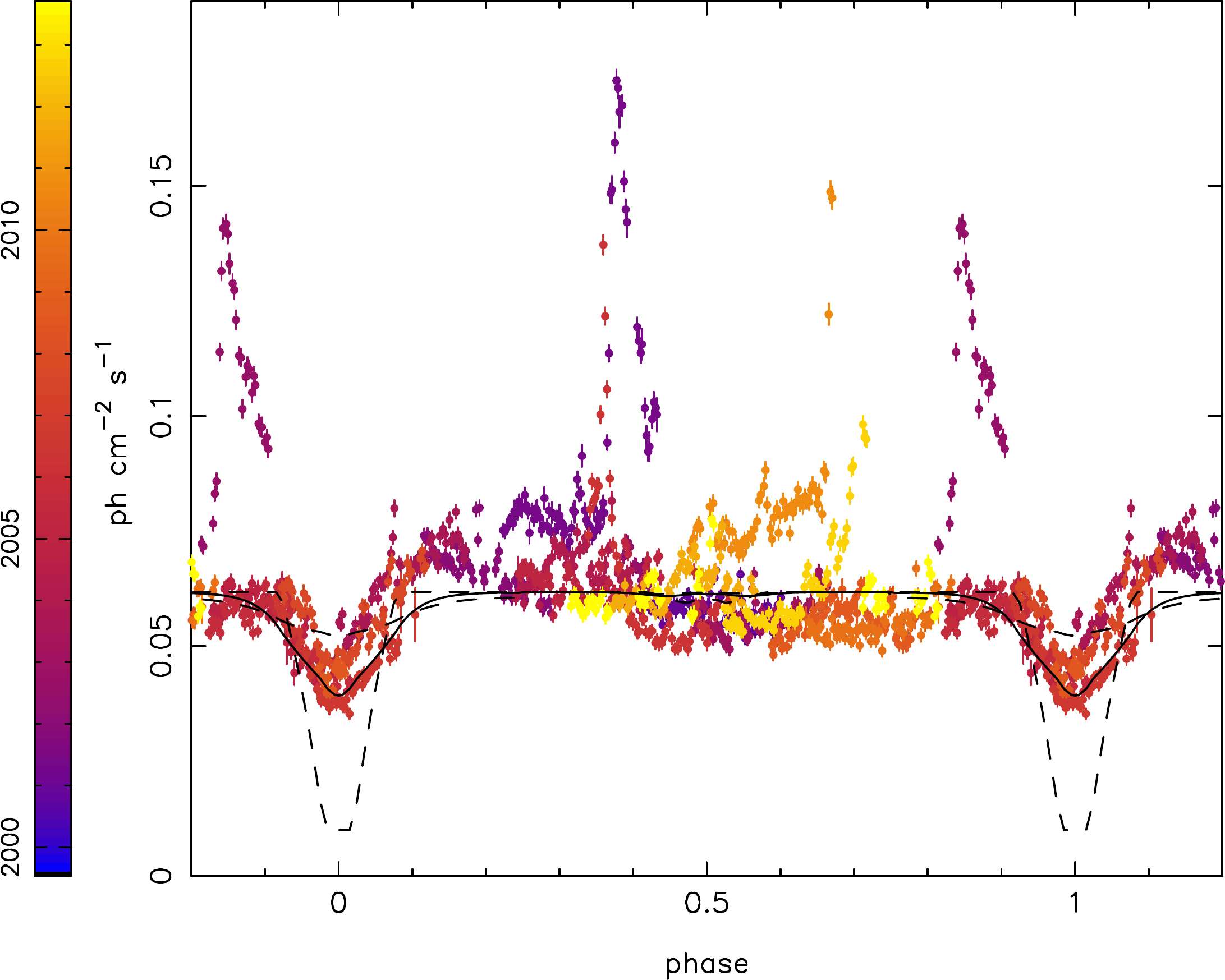}
\end{center}
\caption{Phased composite HRC-I (top) and HRC-S (bottom) X-ray light curves of AR~Lac for all the epochs listed in Table~\ref{t:hrcobs}.  Data are 
colour-coded according
to the time of acquisition and are compared with model light curves for 
spherically-symmetric coronae on both components.  Models, from top to bottom,
are for coronal scale heights of 0.2, 0.5 and 1.0 $R_{\odot}$.  
\label{f:hrcphase}}
\end{figure*}

Data analysis employed standard CIAO v4.5 procedures and calibration database CALDB v4.5.5.  
The event files were partitioned into 400 second bins (and whatever exposure remained in the final bin).  For each bin, spectrum-weighted exposure maps, describing the product of effective area and exposure time, of the detector region under the dithered source region was generated using standard CIAO tools.  The weighting spectrum was computed 
using the APEC model in XSPEC, using four discrete components normalised to match the emission measure distribution as a function of temperature of \citet{Huenemoerder.etal:03}.
We emphasise that the  exact choice of spectral parameters is not important here (see also Section~\ref{s:vstime}).  Net source counts were extracted from circular source regions whose radii depended on off-axis angle, and surrounding background annulus regions.  Energy and photon fluxes were then calculated by dividing the net counts by the exposure map value in the appropriate units of the pixel at the center of the source region. 

A composite of all the \chandra\ HRC-I and HRC-S observations are illustrated as a function of orbital phase in Figure~\ref{f:hrcphase}.

\subsection{EUVE OBSERVATIONS}
\label{s:euve}

AR Lac was observed with the \euve\ Deep Survey (DS) telescope on four separate epochs, one in 1997 July, and three in 2000 September, for a total exposure time of 435~ks.  Details of the observation times are listed in Table~\ref{t:euveobs}. 
Photons gathered by the DS telescope are either intercepted by 
three symmetrically-oriented grazing incidence diffraction gratings, or else pass 
through to the Deep Survey (DS) detector.  On-axis photons, such as would be
observed from a point source during a normal spectroscopic pointing, fall on a boron-coated Lexan filter supported on a nickel mesh, having 
significant transmission between approximately 65 and 190~\AA\ and peaking near 
90~\AA\ with an effective area of about 28~cm$^2$.  A complete description of
the EUVE instrument and its performance can be found in
\cite{Welsh.etal:89} and \cite{Bowyer.Malina:91}.  The sensitivity of the \euve\ instruments in terms of optically-thin plasma emission, as is expected to characterize the coronae of active stars like AR~Lac, has been thoroughly discussed by \citet{Drake:99}.

\begin{deluxetable}{lll}
\tablecaption{EUVE Observations of AR~Lac analysed here \label{t:euveobs}}
\tablehead{
\colhead{UT Start} &
\colhead{UT Stop} &
\colhead{Exposure (s)} 
}
\startdata
1997-07-03 13:40:17 & 1997-07-06 06:25:48 &  84758 \\
2000-09-04 05:46:34 & 2000-09-08 10:12:25 & 137359 \\
2000-09-08 11:08:44 & 2000-09-12 15:31:02 & 132807 \\
2000-09-14 16:47:33 & 2000-09-18 00:56:34 &  79949 \\
\enddata
\end{deluxetable}

\begin{figure}
\begin{center}
\includegraphics[angle=0,width=2.55in]{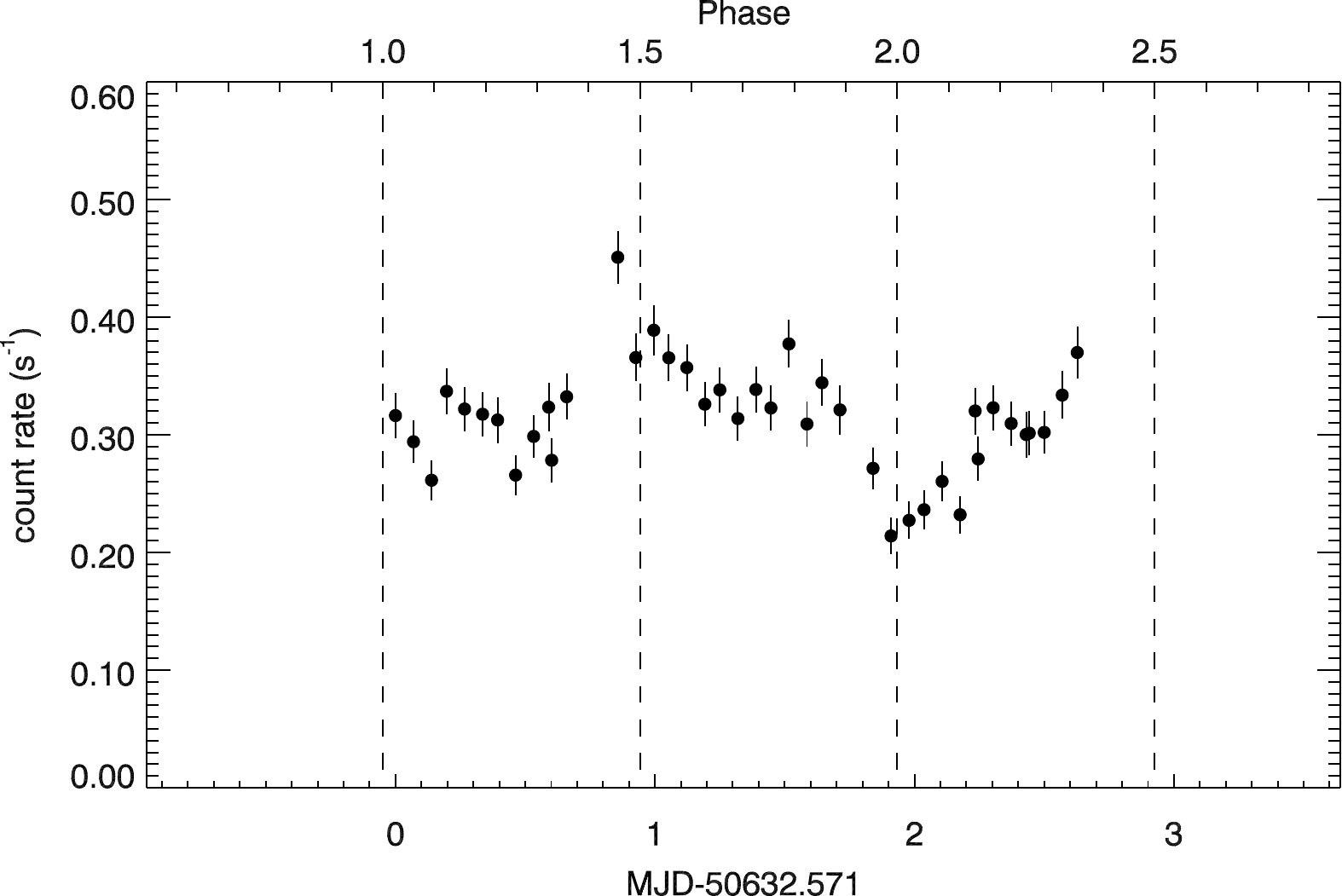}
\includegraphics[angle=0,width=2.55in]{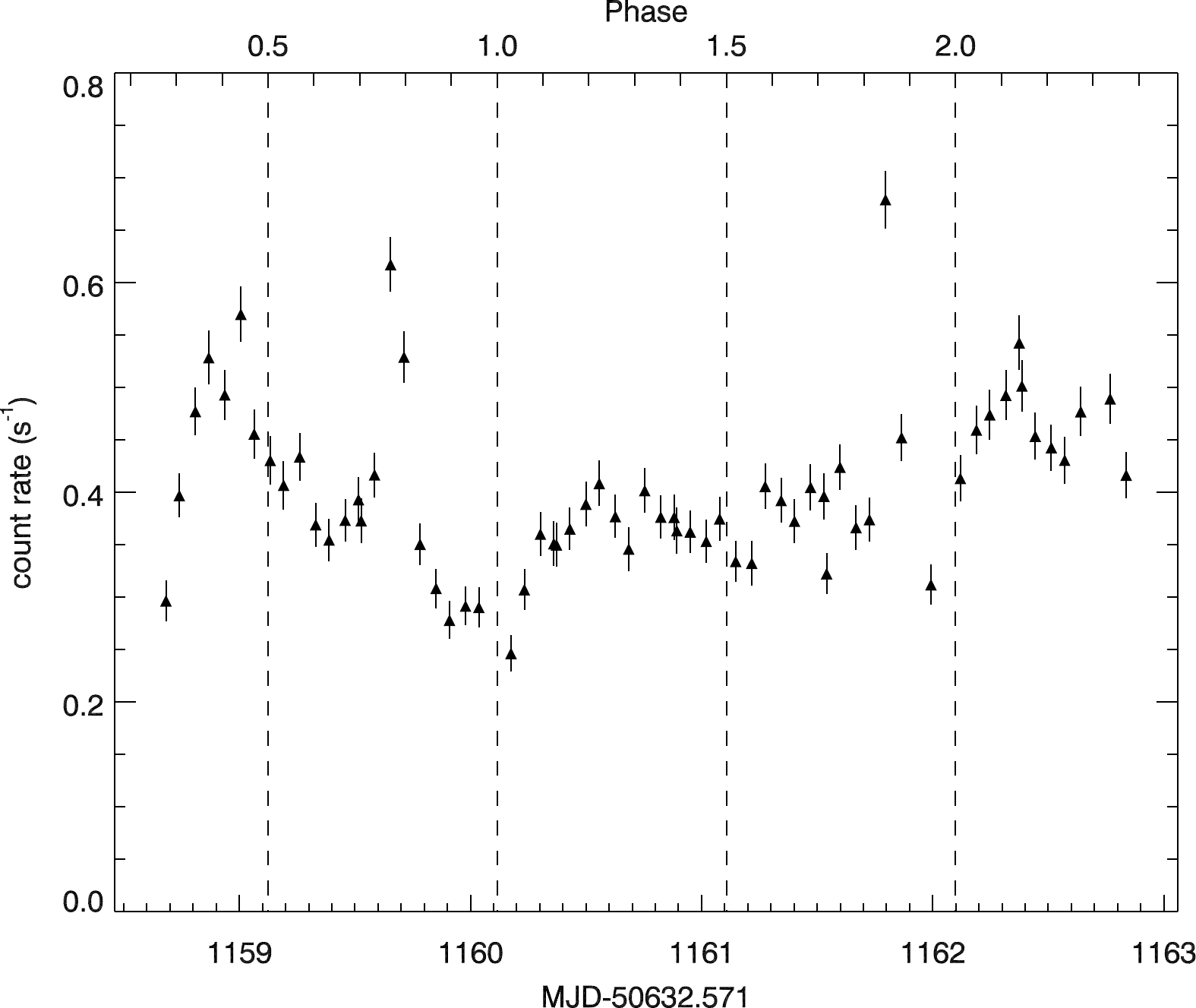}
\includegraphics[angle=0,width=2.55in]{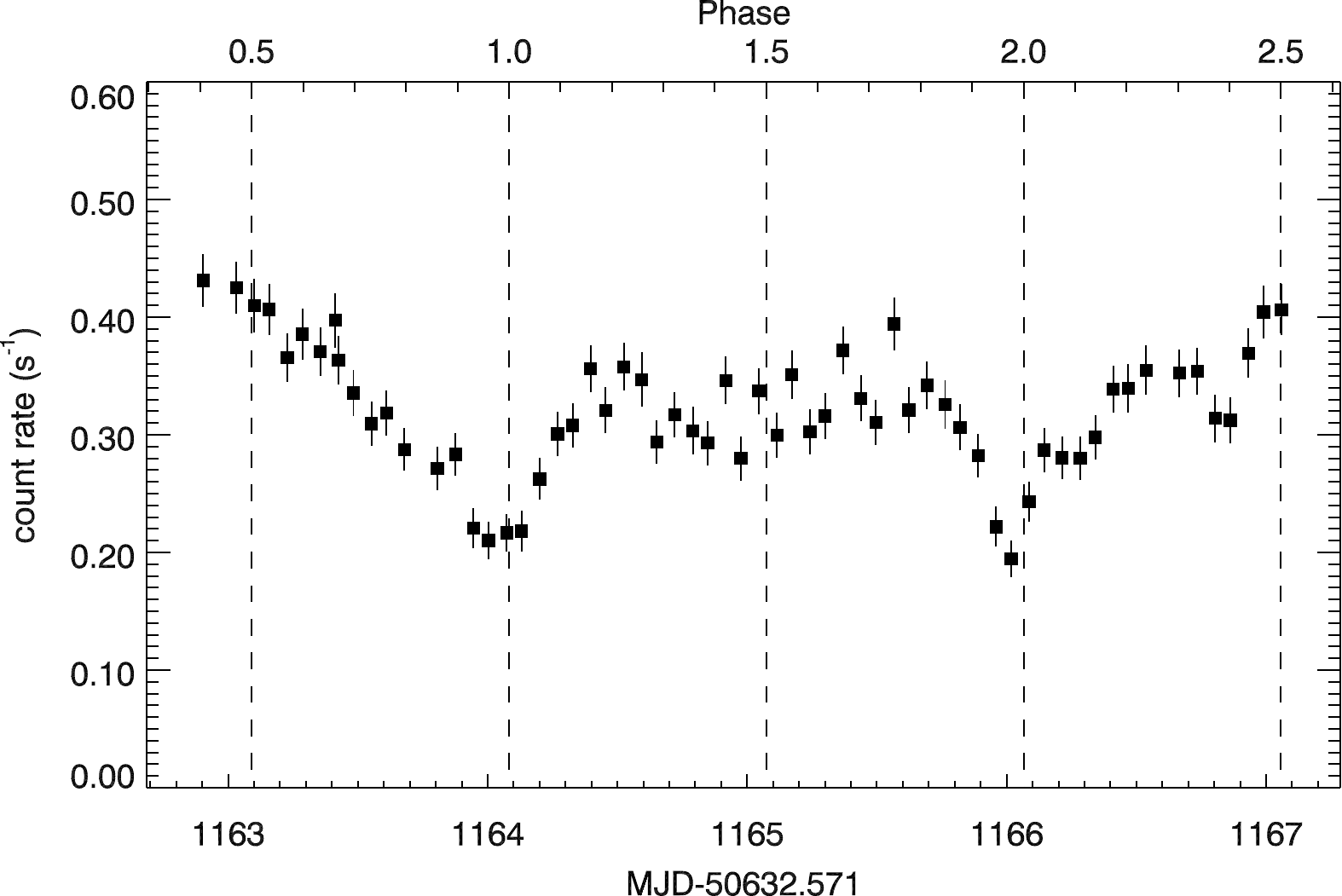}
\includegraphics[angle=0,width=2.55in]{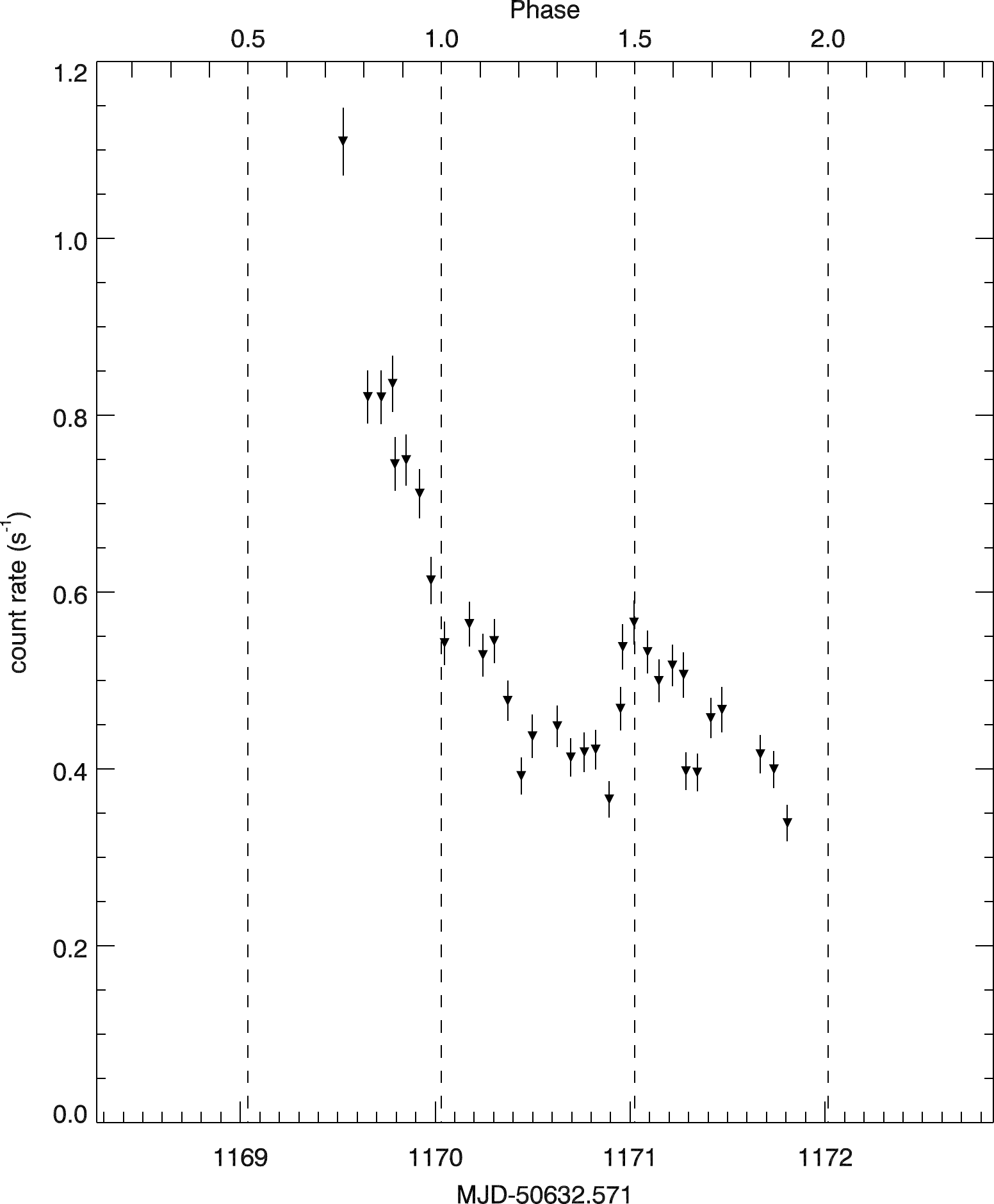}
\end{center}
\caption{
\euve\ light curves for the four epochs analysed here.  Data are binned on 1000~s intervals.  Each panel is sized such that the y-axis scale is the same for all and the bottom panel does not include the flare peak at about 4.6 count~s$^{-1}$; see Figure~\ref{f:euveflare} for the full detail.
\label{f:euveobs}}
\end{figure}

\begin{figure}
\begin{center}
\includegraphics[angle=0,width=3.in]{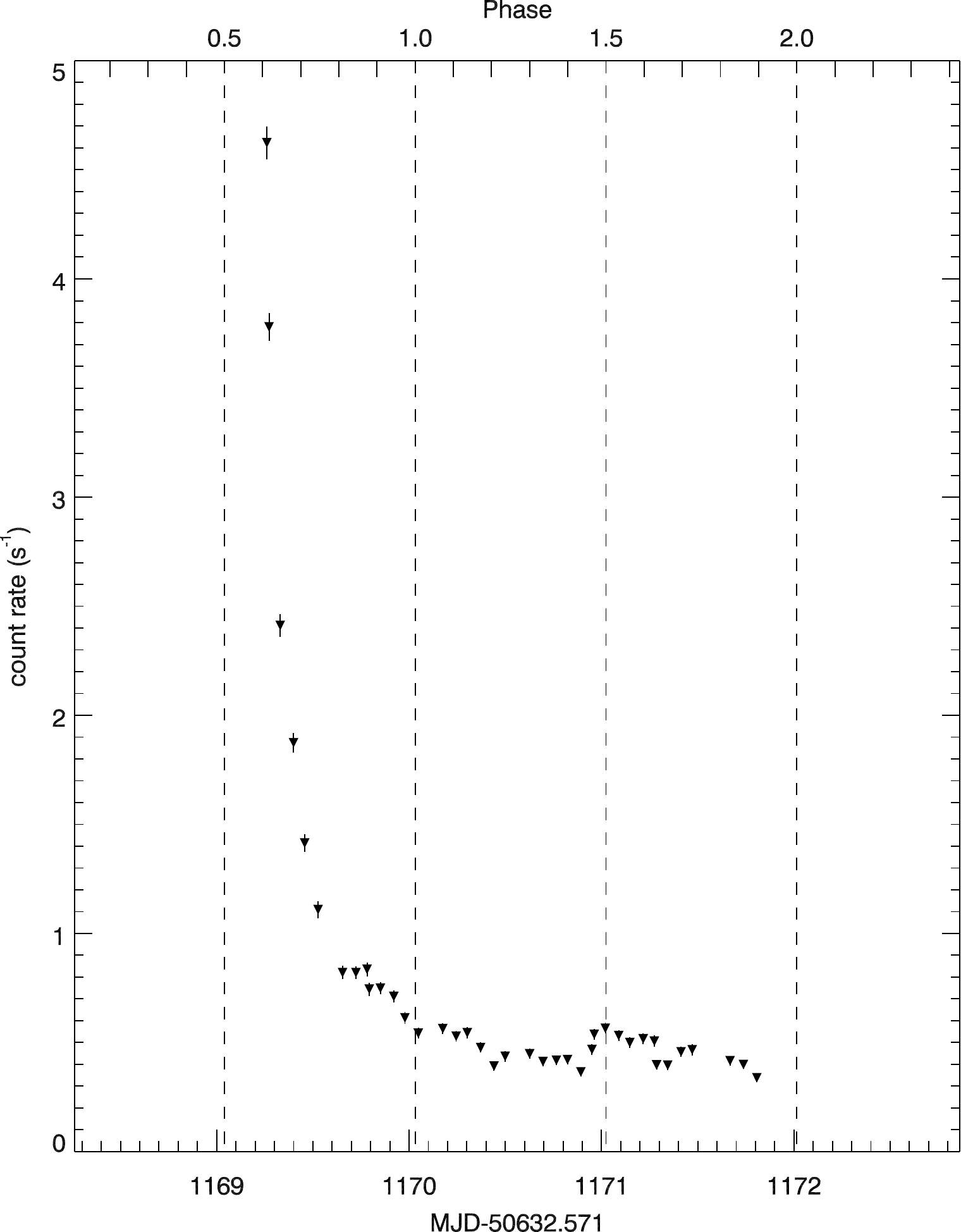}
\end{center}
\caption{Detail of the flare detected by EUVE on 2000 September 14.
\label{f:euveflare}}
\end{figure}

\begin{figure}
\begin{center}
\includegraphics[angle=90,width=3.in]{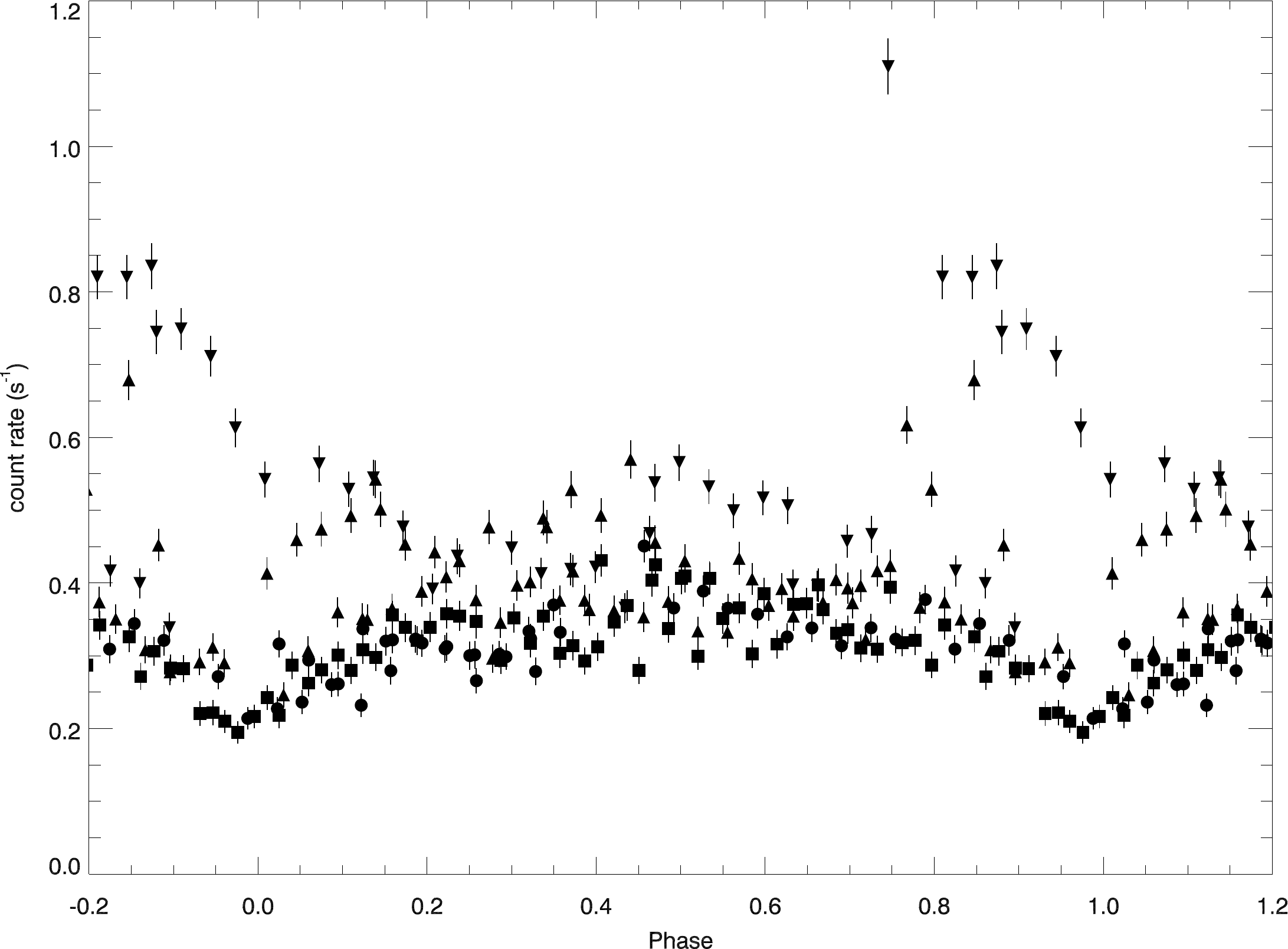}
\end{center}
\caption{\euve\ light curves for all four epochs shown as a function of orbital phase.  Plot symbols correspond to those in Figure~\ref{f:euveobs} for the four epochs analysed here.  Data are binned on 1000~s intervals. 
\label{f:euvephase}}
\end{figure}

We obtained DS QPOE (quick position-ordered event) files from the \euve\ archive and
processed the data using the most current telescope effective 
area, vignetting corrections and ``Primbsh'' corrections for when the telemetry was busy. 
We obtained the DS light curve using the standard \euve\ IRAF software.  Light curves for the separate EUVE observations binned at 1000~s intervals are illustrated in Figure~\ref{f:euveobs}.  A flare amounting to a peak count rate of ten times the quiescent value was detected in the last observational segment; this is illustrated in Figure~\ref{f:euveflare}.  The data are illustrated as a function of orbital phase in Figure~\ref{f:euvephase}.

\section{LIGHT CURVE ANALYSIS}
\label{s:anal}

Both \chandra\ and \euve\ sets of light curves are characterized by stochastic variability and flaring, but also by prominent primary eclipses in the the majority of epochs in which the eclipse was covered.  There are no obvious signs of the secondary eclipse in either the X-ray or EUV data; we will return to this in the discussion below.

\subsection{Constraining the coronal scale height}
\label{s:scaleheight}

The simplest coronal model to consider is a spherically-symmetric shell of emission surrounding both stars that is constant in time.  Such a model has a well-behaved, symmetrical primary eclipse whose width and depth depends on the relative brightness of the two stars and the coronal scale height.  
Unfortunately, Figure~\ref{f:hrceclipses} demonstrates that 
``clean'' and perfectly symmetrical coronal primary eclipses that can easily be interpreted in terms of spherical emitting geometry are fairly rare.   However, there are a number of primary eclipses in which ingress or egress does appear to follow the shape expected for such a spherical shell of emission and that are unaffected by significant flares.  We have extracted these and have fitted a coronal emission model.

\begin{figure}
\begin{center}
\includegraphics[width=3.3in]{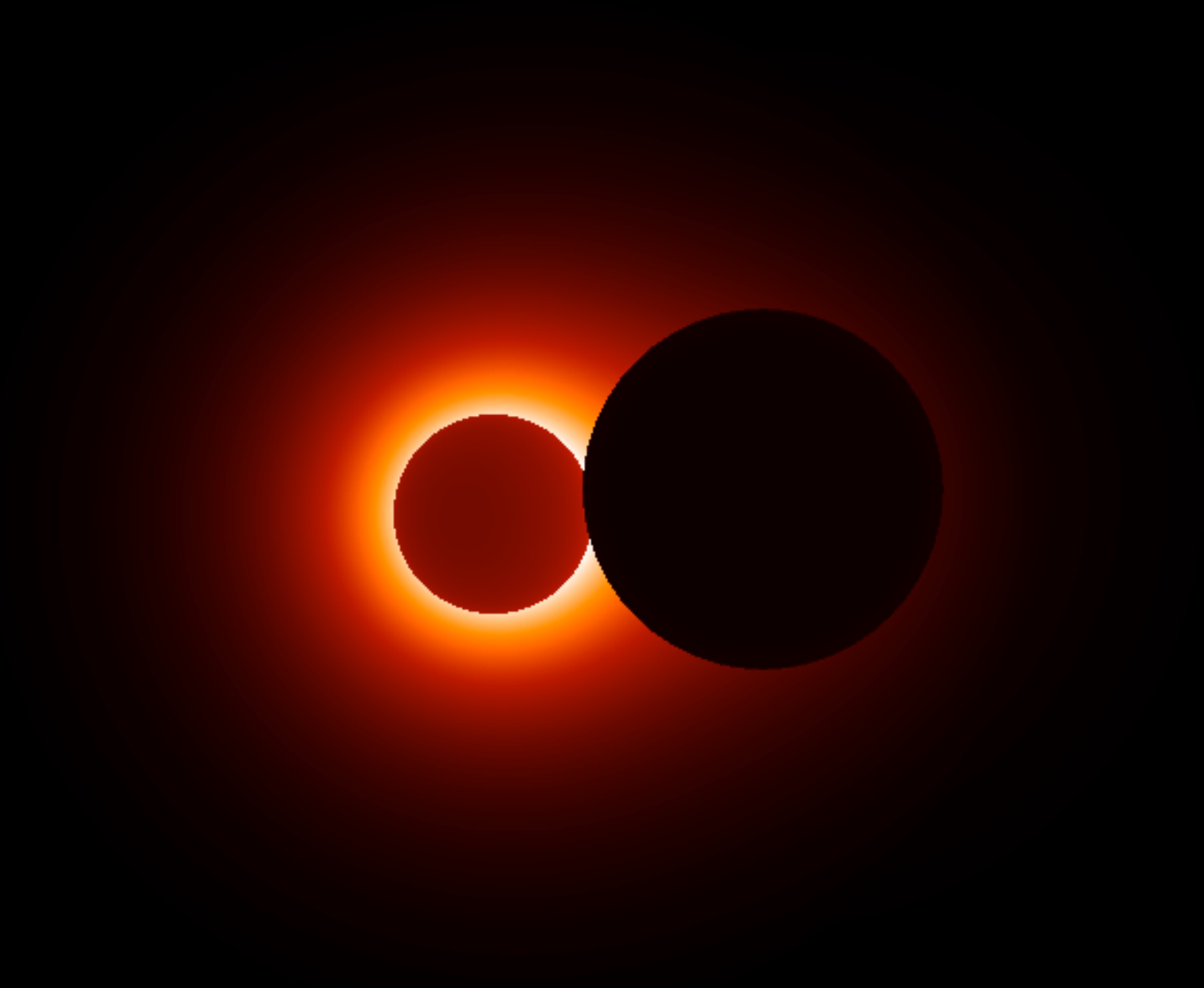}
\end{center}
\caption{Illustration of the spherically-symmetric coronal model as the stars approach primary eclipse with the larger K star in front.  Each star is assumed to be opaque and surrounded by a transparent spherical shell of emission with an exponential decline with height.      The model shown correspond to the parameters $h_K = h_G = 1.3 R_\odot$, and $b_K = 0.44$.
\label{f:modelimage}}
\end{figure}

\begin{figure*}
\begin{center}
\includegraphics[width=6.5in]{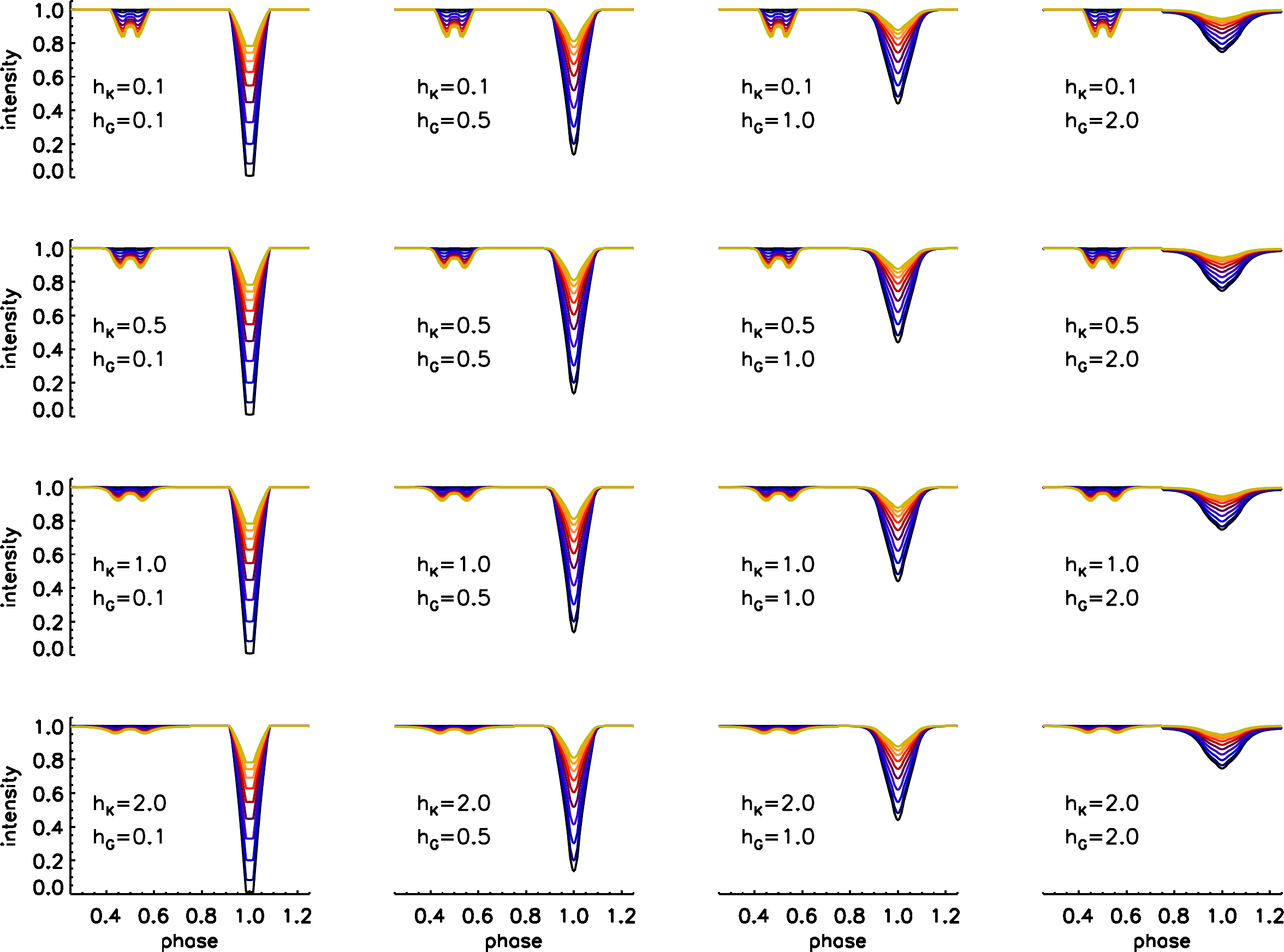}
\end{center}
\caption{Illustration of how the primary and secondary eclipse profiles change with the free parameters in the light curve model: $b_K$, $h_K$ and $h_G$; $b_G$ is normalized to 1 for all models.  The parameter $h_K$ increases from top to bottom and is fixed from left to right.  Conversely, $h_G$ increases from  left to right and is held fixed from top to bottom. The measure of the relative brightness of the primary and secondary coronae, $b_K$ (with $b_G$ being normalised to 1), goes from 0.1 (blue; dark) to 2.0 (yellow; light) in steps of 0.3.
\label{f:modelgrid}}
\end{figure*}

Our model comprises a numerical spherically-symmetric intensity profile exponentially decaying with height placed on each star.  These intensity profiles are placed in a 3D cartesian system and the intensities are projected onto a plane perpendicular to the direction of observation.  The disks of the stars are opaque, and
the atmospheres are assumed to be fully transparent such that neither primary nor 
secondary corona contributes to any dimming of the corona behind it.  
Using the system parameters listed in Table~\ref{t:arlac}, we compute the projected coronal emission as a function of orbital phase and produce artificial light curves.   There are three free parameters: the K star is characterized by a scale height $h_K$ and relative brightness $b_K$, and the secondary G star has a scale height $h_G$ and a fixed brightness $b_G=1$.  The overall normalization of the model is set separately for each dataset to be the median count rate of all rates for a given instrument over the phase range $0.15:0.85$.
These values were found to be 0.074, 0.064, and 0.33 ph~s$^{-1}$~cm$^{-2}$ for {\sl Chandra}/HRC-I, {\sl Chandra}/HRC-S, and EUVE/DS respectively.  An image of illustrating the modelling approach for typical parameters for AR~Lac are illustrated in Figure~\ref{f:modelimage} and 
illustrative model curves for different parameter values are shown in Figure~\ref{f:modelgrid}.

Figure~\ref{f:modelgrid} illustrates the comparative lack of sensitivity of the eclipses to the scale height of the corona of the larger K star, $h_K$, even for the case in which this component is brighter than the G star.  The secondary eclipse depth, when the G star is in front, changes by only 10\%\ or so for K star scale height changes of a factor of 10 or more.  We fail to detect the secondary eclipse unambiguously in any of our observations, and therefore concentrate on the analysis of primary eclipse and the G star coronal scale height.

\begin{deluxetable*}{lcccc}
\tablecaption{Model parameter estimates}
\tablehead{ 
\multicolumn{2}{c}{Dataset} & 
\colhead{$h_{\rm G}$~[R$_\odot$]} & 
\colhead{$b_{\rm K}$} & 
\colhead{$\chi^2/ \nu$}
}
\startdata
Full eclipse & HRC-I & $1.3_{>1.27}^{<1.33}$ & $0.20_{>0.20}^{<0.22}$ & 571/389 \\
Full eclipse & HRC-S & $0.7 _{>0.68}^{<0.74}$ & $1.16_{>1.15}^{<1.16}$ & 344/330 \\
Full eclipse & EUVE & $1.5_{>1.46}^{<1.59}$ & $0.20_{>0.2}^{<0.3}$ & 79/80 \\
Full eclipse & EUVE$+$HRC-I$+$HRC-S & $1.3^{<1.36}_{>1.27}$ & $0.44^{<0.48}_{>0.40}$ & 1357/802 \\
Ingress & EUVE$+$HRC-I$+$HRC-S & $1.25^{<1.33}_{>1.20}$ & $0.53^{<0.60}_{>0.49}$ & 352/358 \\
Egress & EUVE$+$HRC-I$+$HRC-S & $1.4^{<1.44}_{>1.36}$ & $0.29^{<0.38}_{>0.26}$ & 998/366 \\
\enddata
\label{t:bestfit}
\end{deluxetable*}

\begin{figure}
\begin{center}
\includegraphics[width=3.2in]{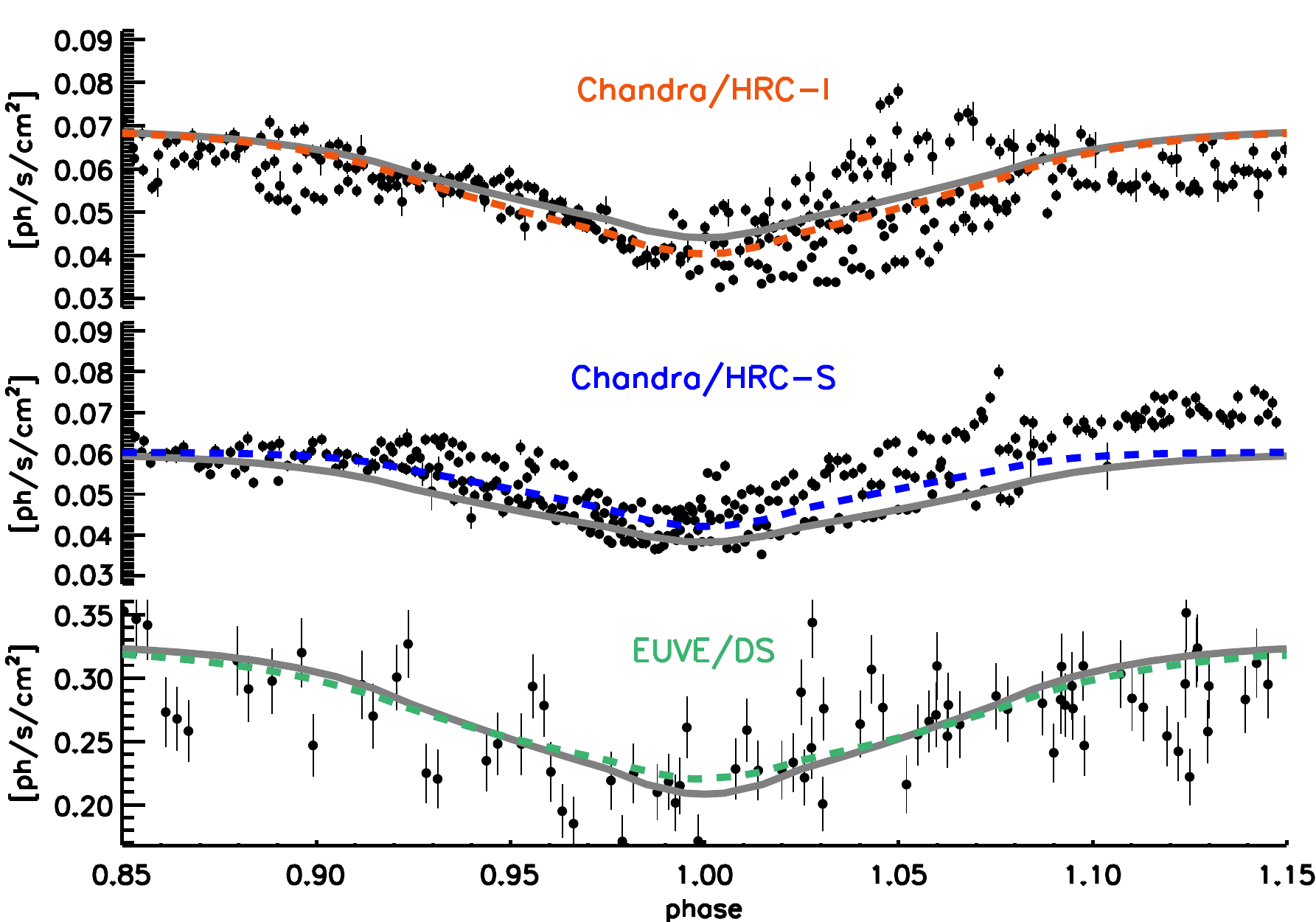}
\end{center}
\caption{Illustration of the best-fit  light curve model overlaid with the \chandra\ HRC and \euve\ data as a function of orbital phase.  Best-fits to all the data combined (grey) and to the individual data sets (colored) are shown.  The best-fit parameters are listed in Table~\ref{t:bestfit}.
\label{f:bestfit}}
\end{figure}

\begin{figure}
\begin{center}
\includegraphics[angle=0,width=3.2in]{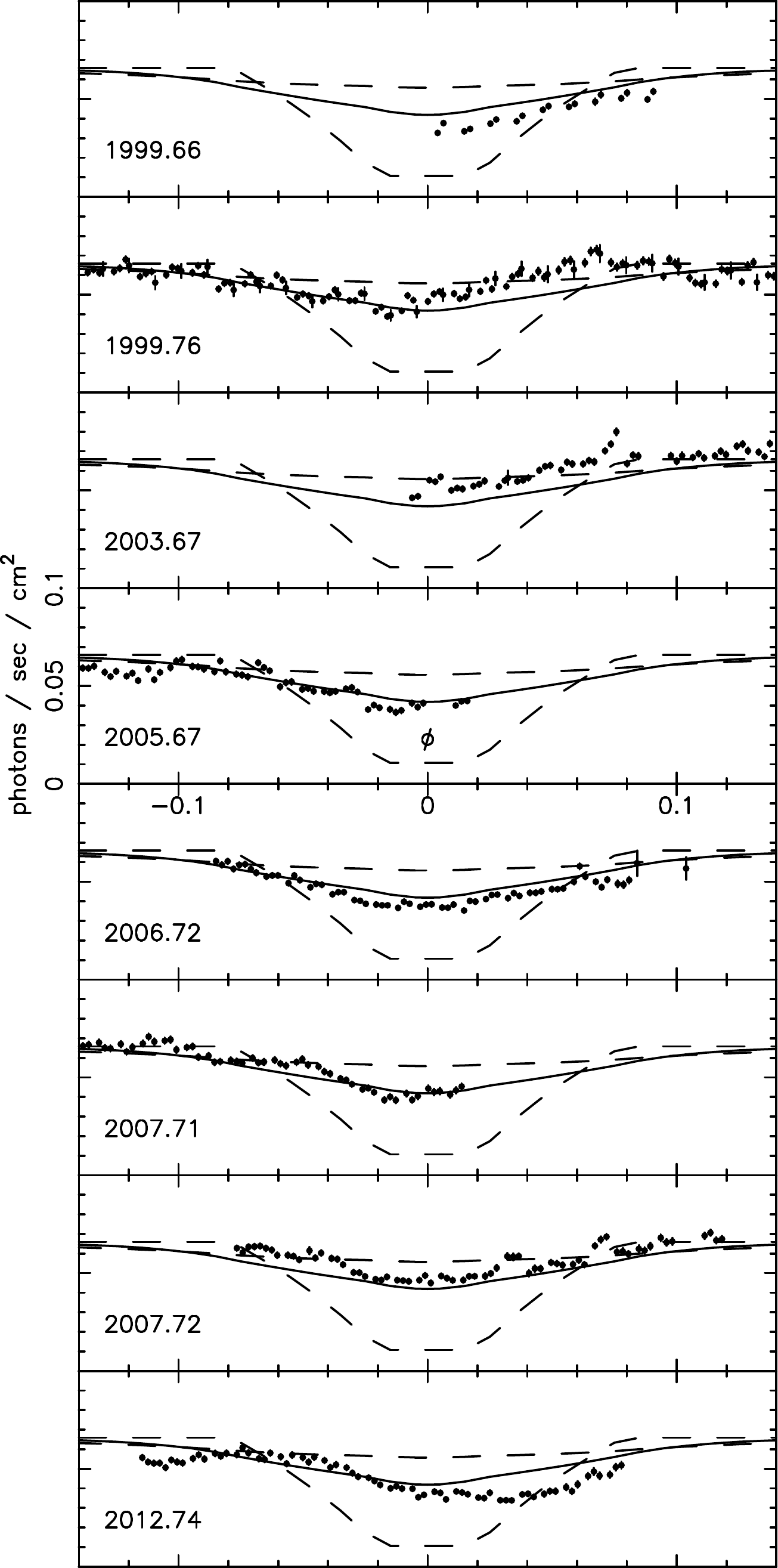}
\end{center}
\caption{\chandra\ HRC X-ray light curves of AR~Lac for data during primary eclipse that are uncontaminated by obvious flaring.  Model light curves for our ``best-fit'' model are also shown 
($h_K=h_G=1.3R_\odot$, $b_K=0.44$; 
see Table~\ref{t:bestfit} for a full summary of fit results), together with those for secondary coronal scale heights of $h_k=h_G=0$  (the deep dashed profile) and $h_K=h_G=2.5 R_\odot$ (the shallow dashed profile).
\label{f:hrceclipses}}
\end{figure}

We performed a brute-force grid search of the model parameters for the 
best-fit to the observations over phases $\pm 0.15$ around the primary eclipse with the K star in front.  This was done for  the different cases of HRC-I, HRC-S and \euve\ data treated both separately, and with all data combined.  For the combined data, fitting was also performed for ingress and egress separately.
The $\chi^2$ values are calculated assuming both a statistical error (${\propto}{\sqrt{{\rm counts}}}$ and a nominal 10\%\ excess error attributable to intrinsic variability.
The likelihood of seeing the observed data, $D$, given the model is then computed as $p(D|h_K,h_G,b_K)=e^{-\chi(h_K,h_G,b_K)^2/2}$.
We adopt flat priors on all parameters over the grid range, and multiply them with the likelihood.
Following Bayes' Theorem, this generates the joint posterior probability density distribution of the parameters given the data,
\begin{equation}
p(h_K,h_G,b_K|D) \propto p(h_K) p(h_G) p(b_K) p(D|h_K,h_G,b_K)
\end{equation}
We then obtain posterior density distributions for each parameter by marginalizing over the other two,
\begin{eqnarray}
p(h_K|D) &\propto& \int d\,h_G d\,b_K p(h_K,h_G,b_K|D) \\
p(h_G|D) &\propto& \int d\,h_K d\,b_K p(h_K,h_G,b_K|D) \\
p(b_K|D) &\propto& \int d\,h_K d\,h_G p(h_K,h_G,b_K|D), 
\end{eqnarray}
although in practice we have no information to usefully constrain the scale height of the K star corona, $h_K$.
The modes of the distribution correspond to locations of minimum $\chi^2$.
These, and the 68\%\ half-tail credible regions are reported in Table~\ref{t:bestfit}.
The model curves for the best-fit values obtained by jointly fitting all datasets (excluding periods of obvious flares) are shown in Figure~\ref{f:bestfit}.

The best-fit model parameters generally indicate that the G star corona has a scale height of about 1.3 solar radii and is brighter than that of the K star by a factor of 2--5. 
The fit parameters for HRC-I and \euve\ are reasonably consistent with one another, while the best-fit HRC-S scale height is a factor of two smaller and the relative brightness parameter, $b_K\sim 1$, suggests both stars are of approximately equal brightness.  However, visual inspection of the data and models in Figure~\ref{f:bestfit} reveal substantial deviations between them, particularly at egress.  This will be discussed further in Sect.~\ref{s:discuss} below.

The results for ingress and egress treated separately are generally consistent for the values of coronal scale height, within the uncertainties of the measurements.  The relative brightness parameter is marginally different though, with a best-fit value slightly larger by about 40\%\ for ingress than egress.  

\subsection{Flares}
\label{s:flares}

Both \chandra\ and \euve\ observations are characterized by a number
of flares.  The \chandra\ flares could all be described as fairly
modest, with the largest having peaks only a factor of 3 or so higher
than the quiescent level, and decay timescales of a few ks. A full
analysis of this flaring component is beyond the scope of this paper.
The last \euve\ epoch that began on 2000 September 14, however, happened upon a large event whose rise
phase occurred before the observation start, but whose decay was
tracked over two days (see Figure~\ref{f:euveflare}.  The spectrum of the event 
was previously analyzed by \citet{Sanz-Forcada.etal:03}
who reported a flare fluence of $2.0\times10^{35}\,\mathrm{ergs}$.

To refine the flare characteristics, we fitted the \euve\ lightcurve
of this large flare using Weibull distributions following the method
described in detail by \citet{Huenemoerder.etal:10}.  This normalized
distribution is defined by the equations
\begin{equation}
  f(p,a,s)=\left( \frac{a}{s} \right) p^{(a-1)} e^{-p^a}
\end{equation}
\begin{equation}
  p=(t-t_0)/s
\end{equation}
where $a$ is a shape parameter and takes values $a > 0$, $s$ describes
the scale or width of the distribution and is also positive, $s > 0$,
and $t$ is the time coordinate with the time of flare onset given by
$t_0$.  An amplitude parameter normalizes to the total counts, and we
included a constant term as an estimate of the quiescent rate.  While
this is an empirical parameterization of a flare, the function
can range from purely exponential form for $a=1$, to steeper for $a<1$, or
shallower for $a>$, to smooth rise and decay for large $a$.

\begin{figure}
\begin{center}
\includegraphics[angle=90,width=3.3in]{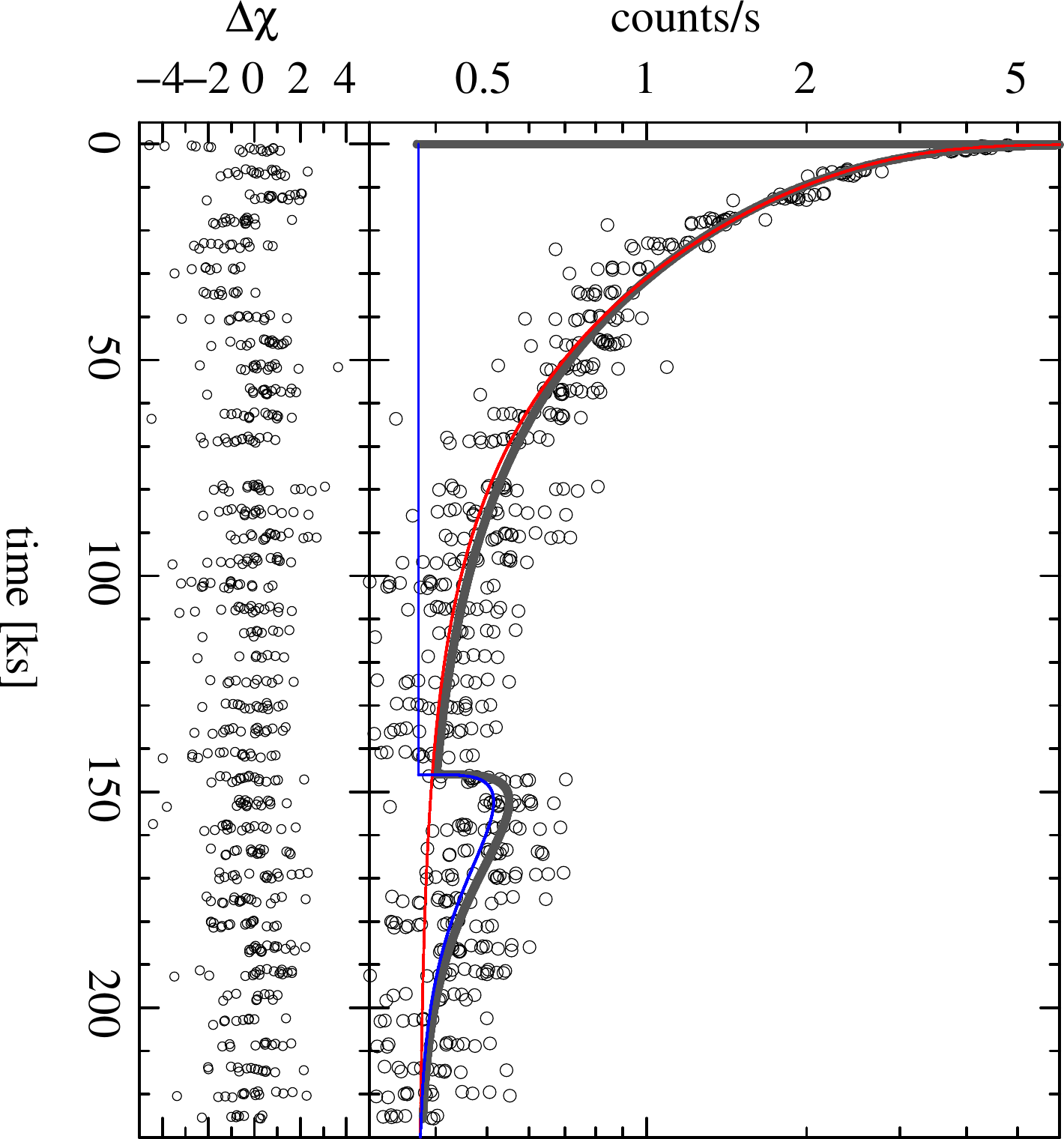}
\end{center}
\caption{\euve\ DS light curve, binned on 100s intervals, and best-fit Weibull distribution model and residuals for the large flare of 2000 September 14--18.  The grey curve represents the overall model, while the red and blue curves illustrate the models for the large and smaller flared, respectively.  
\label{f:euvflaremodel}}
\end{figure}

We fitted both the large flare, and a smaller event that occurred nearly
two days after the observation start.  The free parameters in the
model are the normalization factor, $a$, $s$, and $t_0$ and the
best-fit is illustrated in Figure~\ref{f:euvflaremodel}.  The large
flare contains 66000 counts, a scale of $25\,\mathrm{ks}$, and is
slightly steeper than exponential ($a=0.82$), while there are 4700
counts in the smaller flare, offset by $146\,\mathrm{ks}$, the same
scale, but slightly slower than exponential ($a = 1.2$).  The constant
term had a value of $0.37\,\mathrm{counts\,s^{-1}}$.

To estimate the flare energy, we used two methods.   First, the scaling relations between observed counts and source flux for an isothermal spectrum with a flare-like temperature of $2\times 10^7$~K given by \citet[][Table 8]{Drake:99} indicate that one count in the \euve\ DS Lex/B filter corresponds to an energy of about $3\times 10^{-11} (2\pi D^2)$~erg, 
in the 65--195~\AA\ bandpass, for a distance $D$. Here, we used an interstellar absorption column density of $N_\mathrm{H}=2\times10^{18}\,\mathrm{cm^{-2}}$ derived by
\citet{Walter:96} based on \euve\ spectra of AR~Lac, which is similar
to the value derived by \citet{Sanz-Forcada.etal:03}.  Since a flare is thought to arise on a fairly discrete, limited part of the corona, we have also divided the flux from \citet{Drake:99} by two to remove the correction factor introduced to account for the ``unseen" corona on the far side of the star \citep[see, e.g.,][]{Jordan.etal:87}.  For a total of 66000 counts, we find an EUV fluence of $2.1\times 10^{35}$~erg.

Secondly, we adopted the plasma model (emission
measure distribution (EMD) and elemental abundances) of
\citet{Huenemoerder.etal:03} derived from HETG spectra obtained around the same time as the \euve\ data, between 2000 September 11--19.  However, we found their normalization too
high, and had to rescale the model to an integrated emission measure
of $4.9\times10^{53}\,\mathrm{cm^{-3}}$ to better match the X-ray
spectrum of the approximately steady flux of Chandra Observation ID 9.
This EMD also closely matches that derived by
\citet{Sanz-Forcada.etal:03} from the \euve\ flare
spectrum.  As such, it represents some mean of flare and quiescent
plasma EMDs.  

For this model and a distance of $42\,\mathrm{pc}$, the EUV flare photons
(in the range of $65$--$190\,$\AA) represent a fluence of
$1.8\times10^{35}\,\mathrm{ergs}$ (a lower limit since the flare onset
was not observed), in good agreement with both \citet{Sanz-Forcada.etal:03} and the estimate based on the \citet{Drake:99} tables. Scaling this to the HETG band
($1.7$--$25\,$\AA) would produce an X-ray flare fluence of
$1.3\times10^{36}\,\mathrm{ergs}$, indicating that the radiative output of the hot flare plasma is dominated by X-ray emission.

While the flare looks very large in the EUV, the estimated X-ray flare
fluence is typical for large flares of young active stars, which have a similar activity level as RS~CVn binaries \citep[see, e.g.][figure 7]{Schulz.etal:06}.

\section{DISCUSSION}
\label{s:discuss}

Eclipse observations provide potentially powerful diagnostics of the geometry of the emitting regions and have provided the main motivation for studies of the outer atmospheres and coronae of the AR~Lac system.  Figures~\ref{f:hrcphase} and \ref{f:euvephase}, however, starkly illustrate the difficulties in interpreting such data in terms of obscuration and rotational modulation.  The X-ray and EUV source flux is in a state of frequent change on a variety of timescales, and it is often unclear whether these variations are due to geometrical effects or simply reflect stochastic brightening and dimming of the emitting regions.  Clear rotationally-modulated variations in coronal EUV and X-ray emission of active stars have often been sought after and occasionally seen but are not common and even when identified tend to account only for a fraction of the observed variations \citep[see, e.g.][]{Agrawal.Vaidya:88,Drake.etal:94,Guedel.etal:95,Guedel.etal:95b,Kuerster.etal:97,Audard.etal:01,Garcia-Alvarez.etal:03,Marino.etal:03,Flaccomio.etal:05}.  Attempts at geometrical reconstruction of the emitting regions based on limited coverage, or only a single rotation phase, in which all variations are assumed to arise from rotational modulation, are then very likely to result in spurious structure.

\subsection{Coronal scale height and substructure}

The light curve modelling described in Sect.~\ref{s:scaleheight} at face value succeeded in constraining the coronal scale height of the G-type primary but not that of the K-type secondary star.  There are three issues to consider in interpreting the results.  Firstly, while it appears that simple, spherically-symmetric models with only 3 free parameters are capable of providing a reasonably good match to the observed light curves, a quasi-infinite range of models of increased complexity and asymmetry could be constructed that could match the observations equally well (or hopefully better, given more free parameters).
Secondly, the observed eclipses do not all have the same profile, and can differ quite significantly between epochs.  This is evident from the \euve\ phased light curve illustrated in Figure~\ref{f:euvephase}, but is graphically illustrated in Figure~\ref{f:hrceclipses}, showing sequentially the \chandra\ eclipses that are unaffected by flaring.   While some portions of the data follow the best-fit eclipse model quite well, such as epochs 2005.67 and the first half of the eclipse of epoch 1999.76, this is the exception rather than the norm.   Thirdly, the model assumes that the emission is effectively constant throughout the eclipse, an assumption that is rendered catastrophically inappropriate during flares, but that also might be questioned during relative quiescence.   We can assess the latter to some extent by examination of the light curves out of eclipse near quadrature phases, when only rotational modulation is otherwise at work.  

The \euve\ observations covering more than one orbital phase in Figure~\ref{f:euveobs} demonstrate that brightness as a function of orbital phase is not very repeatable.  On shorter timescales, 
the non-flaring data are mixed in terms of variability. There are periods of stability with little variation, such as in the HRC-S data illustrated in Figure~\ref{f:hrcphase} between phases 0.6 and 0.9, yet there are also a lot of brightenings and dimmings at other times, such as in the HRC-I data at phases 0.1-0.4.  
Asymmetries in Mg~II line profiles associated with active regions on the K component were detected by \citet{Pagano.etal:01}, and it is likely that some of the eclipse asymmetries and other secular variations on orbital timescales observed here are a coronal signature of analogous active regions.  \citet{Pagano.etal:01} also noted, however, that emission on the G component appeared more uniform.  Small variations might also be caused by absorbing material in the line-of-sight, as suggested by  \citet{Walter:96} based on EUV observations and inferred from the UV study of \citet{Pagano.etal:01}.  In the former case, \citet{Walter:96} estimated an equivalent absorbing column, $n_H$, of only $10^{19}$~cm$^{-2}$ would be required, which would likely not have a noticeable affect on the higher energy X-rays observed in this study.
 
There is a limit to how much the non-flare variations can be caused by simple rotational modulation while still providing eclipses that follow to a reasonable degree what would be expected from a spherically-distributed corona.  A compact bright active region hoving into view around the limb, for example, can cause a fairly rapid brightening, but would also cause a sharp drop  as it was obscured during an eclipse.  It seems more likely that the most rapid variations observed are largely changes in brightness of visible regions in the coronae of the stars than rotational modulation.  


By the same argument, asymmetries in the observed eclipse profiles betray either a change in brightness of the uneclipsed plasma, spatial inhomogeneity in coronal emission, optically-thick absorbing material, or a mixture of these.  Based on the eclipses observed by the \chandra\ HRC, we can point to the following epochs that have essentially full eclipse coverage but asymmetric eclipse profiles: 1999.76, whose eclipse appears narrow with an early egress; 2007.72, whose eclipse also appears narrower than spherical model predictions with a late ingress; and 2012.74, whose eclipse appears broadened with egress shifted to later times.  

Narrow eclipses point to less emission from the stellar limb and might be associated with intrabinary emission located between the two stars, as has been inferred is the case in previous work.  \citet{Siarkowski.etal:96} applied an iterative deconvolution technique described by \citet{Siarkowski:92} to \asca\ observations in 1993 June covering slightly more than one orbital cycle.  Unlike the observations presented here, secondary eclipse did appear to be present in those data, and the reconstruction of \citet{Siarkowski.etal:96} concluded that the emission was dominated by coronal structures located between the two binary stars.  We again note that this reconstruction was based on only one orbital phase, however, and it is likely that the details of the deduced spatial extent of the emission are a spurious manifestation of sort of time-variable emission afflicting the {\it Chandra} data as discussed above.  We therefore temper interpretation of this reconstruction with some degree of caution, although it is also notable that in this case there is an element of support for the deduced intrabinary emission from optical surface features and from Mg~II line profiles of enhanced Mg II emission likely associated with extended structures co-rotating with the K~star and close to the system center-of-mass \citep{Pagano.etal:01}.

It is likely that coronal emission is associated with surface spots found from photometric modulation.  \citet{Lanza.etal:98} found that the spatial association between photospheric spots revealed by optical photometry and chromospheric and coronal plages as detected in the UV \citep[e.g.][]{Pagano.etal:01} and the intrabinary emission deduced in X-rays by \citet{Siarkowski.etal:96} is significant for a large active region inferred around the substellar point on the secondary and is suggested also for smaller starspots on both components.  The large spots on RS~CVn-type binaries, including AR~Lac, appear to be fairly stable, but migrate in phase over time.   On AR~Lac, they are only easily discernible on the larger and optically brighter K seconary star from optical photometry and have been modelled by two large spots
\citet[e.g.][]{Rodono.etal:86,Lanza.etal:98,Siviero.etal:06}.  The migration rate of the spots was found to be 0.4~period~yr$^{-1}$ for data obtained in the years 1978--1981 \citep{Rodono.etal:86} and 0.55~period~yr$^{-1}$ for 2000--2005.  Other spot activity on top of this pattern appears to be more irregular, with \citet{Siviero.etal:06} noting that the light curve shaped by spots does not repeat cycle after cycle.  Based on this spot behaviour, we would not expect the X-ray morphology to be stable over long periods of time, and it is perhaps not surprising that we do not find evidence for the strong intrabinary emission that appeared to characterise the 1993 \asca\ observations, especially if such emission depends on spot activity on both stars being concentrated on the opposing hemispheres.


Proceeding with the assumption that our simple three-parameter spherically-symmetric models give a reasonable average approximation to what is more likely a distribution of discrete regions of varying brightness over the stars, we find the G star  coronal scale height from Sect.~\ref{s:scaleheight} to be about $1.3 R_\odot$, or in terms of the G star radius, about $0.86 R_\star$.  This is considerably larger than the coronal scale height on the Sun, whose typical loop lengths extend to heights of $0.04-0.4 R_\odot$ \citep[e.g.][]{Aschwanden:11}.  This observed loop scale height also corresponds to the pressure scale height $h_\odot=2kT_e/\mu m_H g$: for an active solar coronal temperature of typically about $T_e=2\times 10^6$~K and with $m_H$ being the proton mass, $h_\odot=0.2 R_\odot$.  A larger scale height for the AR~Lac coronae is expected na\"ively because of lower surface gravities and higher coronal temperatures.  For the G primary, the surface gravity is about half the solar value, while the typical coronal temperature is $T_e\sim 10^7$~K, or five times higher than the solar corona.  The scale height is then about ten times larger or $\sim 2 R_\odot$.  This is approximately compatible with our observations.  The surface gravity of the K component is a factor of 3 lower still, and the hydrostatic scale height is about $6 R_\odot$.  

The coronal scale height we find from direct geometrical eclipses is considerably larger than the $0.05 R_\star$ height inferred from {\it EUVE} spectra for ``hot'' loops with temperatures in excess of $10^7$~K by \citep[][see also \citealt{Griffiths.Jordan:98}]{Griffiths:99}, and only slightly more consistent with their finding of $0.15 R_\star$ for cooler loops with temperatures less than $10^7$~K.  Their estimates are based on energy balance models that are somewhat dependent on the adopted gas pressure.  \citet{Pagano.etal:01} found that eclipses of the K component by the G star in the light of Mg~II lines were wider than the simple photospheric geometrical prediction, pointing to a significant extension of the Mg~II emitting gas above the stellar surface.  Our failure to detect the eclipse of the K star in X-rays points to an extended corona, as might be expected from the greater scale height, although formally we cannot provide firm observational constraints on this.

More direct measures of coronal scale height on RS~CVn-type binaries are difficult to obtain.  \citet{Testa.etal:04b} used a detection of resonance scattering in lines of O~VIII and Ne~IX in the coronae of  II~Peg and IM~Peg to infer small scale heights of only a few percent or less of the stellar radius.  Since such scattering was not common among the spectra they investigated, it is possible that during those particular observations the visible hemisphere emission was dominated by a bright active region core.  Flare scale heights based on Fe~K$\alpha$ photospheric fluorescence emission of $\la 0.15R_\star$ ($0.5R_\odot$) on II~Peg and $\la 0.3R_\star$ ($4R_\odot$) on the active K giant HR~9024  by  \citet{Ercolano.etal:08} and \citet{Testa.etal:08}, respectively, are more consistent with our geometric coronal heights.  Eclipsed flares have been observed twice on Algol (B8~V+K2~III), whose optical secondary is similar to the evolved components of RS~CVn-type binaries.  The inferred heights of flaring loops are $\la 0.6R_\star$ ($2.1R_\odot$) \citep{Schmitt.Favata:99}  and $\sim 0.1R_\star$ ($0.35R_\odot$) \citep{Schmitt.etal:03}; see also \citet{Sanz-Forcada.etal:07}.  Our inference of scale height on AR~Lac is then similar to other geometrical results for similar  RS~CVn-like stars. 

\section{X-rays from AR Lac through time}
\label{s:vstime}

\begin{deluxetable}{llcl}
\tablecaption{Multi-decade Observation Information}
\tablehead{ 
\colhead{} & \colhead{}  & \colhead{Rate\tablenotemark{$\dagger$}}   & \colhead{Flux\tablenotemark{$\ddagger$} }     \\   
\colhead{Instrument}	& \colhead{Date} & \colhead{(cts s$^{-1}$)} & \colhead{(erg cm$^{-2}$ s$^{-1}$)}}

\startdata
\einstein\ IPC	& 1980 Jun 14	& $1.9 \pm 0.2\tablenotemark{a}	$		& $3.2\times10^{-11}$	\\
\exosat\ LE			& 1984 Jul 4	& $0.2 \pm 0.02$\tablenotemark{b}	& $2.5\times10^{-11}$		\\
\exosat\ ME			& 1984 Jul 4	& $0.5\pm 0.15$\tablenotemark{b}			& $2.4\times10^{-11}$	\\
\rosat\ PSPC		& 1990 Jun 18 & 	$ 5.0 \pm 0.5$\tablenotemark{c} & $3.2\times10^{-11}$	\\
\rosat\ PSPC			& 1990 Dec 11 & $6.0\pm 1.0$\tablenotemark{d}		& $3.9\times10^{-11}$	\\
\asca\ SIS		& 1993 Jun 2	& $1.05 \pm 0.15$\tablenotemark{e}	& $2.0\times10^{-11}$	\\
{\it Beppo-SAX} & 1997 Nov 2,9 & \nodata & $2.8\times 10^{-11}$\tablenotemark{f} \\
\chandra\ HETG		& 2000 Sep 11	& $1.0 \pm 0.05$\tablenotemark{g}	& $3.4\times10^{-11}$\tablenotemark{g}  	\\
\chandra\ HETG		& 2000 Sep 17	& $1.0 \pm 0.05$\tablenotemark{g}	& $3.8\times10^{-11}$\tablenotemark{g}  	\\
\enddata
\tablenotetext{$\dagger$}{Count rates listed are our own assessments of the quiescent rate outside of eclipses based on data presented in the references indicated.}
\tablenotetext{$\ddagger$}{Except where noted, fluxes are based on the spectral model described in Sect.~\ref{s:chandra} and refer to the 0.5--5.0~keV bandpass.}
\tablenotetext{a}{\citet{Walter.etal:83}}
\tablenotetext{b}{\citet{White.etal:90}; count rate refers to LE1 only}
\tablenotetext{c}{\citet{Ottmann.etal:93}}
\tablenotetext{d}{\citet{Schmitt:92}}
\tablenotetext{e}{\citet{White.etal:94}; count rate refers to a single SIS}
\tablenotetext{f}{\citet{Rodono.etal:99}; flux taken directly from their Table 2 with uncertainty based on light curve variations in their Fig.~2.}
\tablenotetext{g}{\citet{Huenemoerder.etal:03}; flux determined by direct integration of HETG spectrum}

\label{t:xrayvstime}
\end{deluxetable}

\begin{figure*}
\begin{center}
\includegraphics[angle=0,width=6.5in]{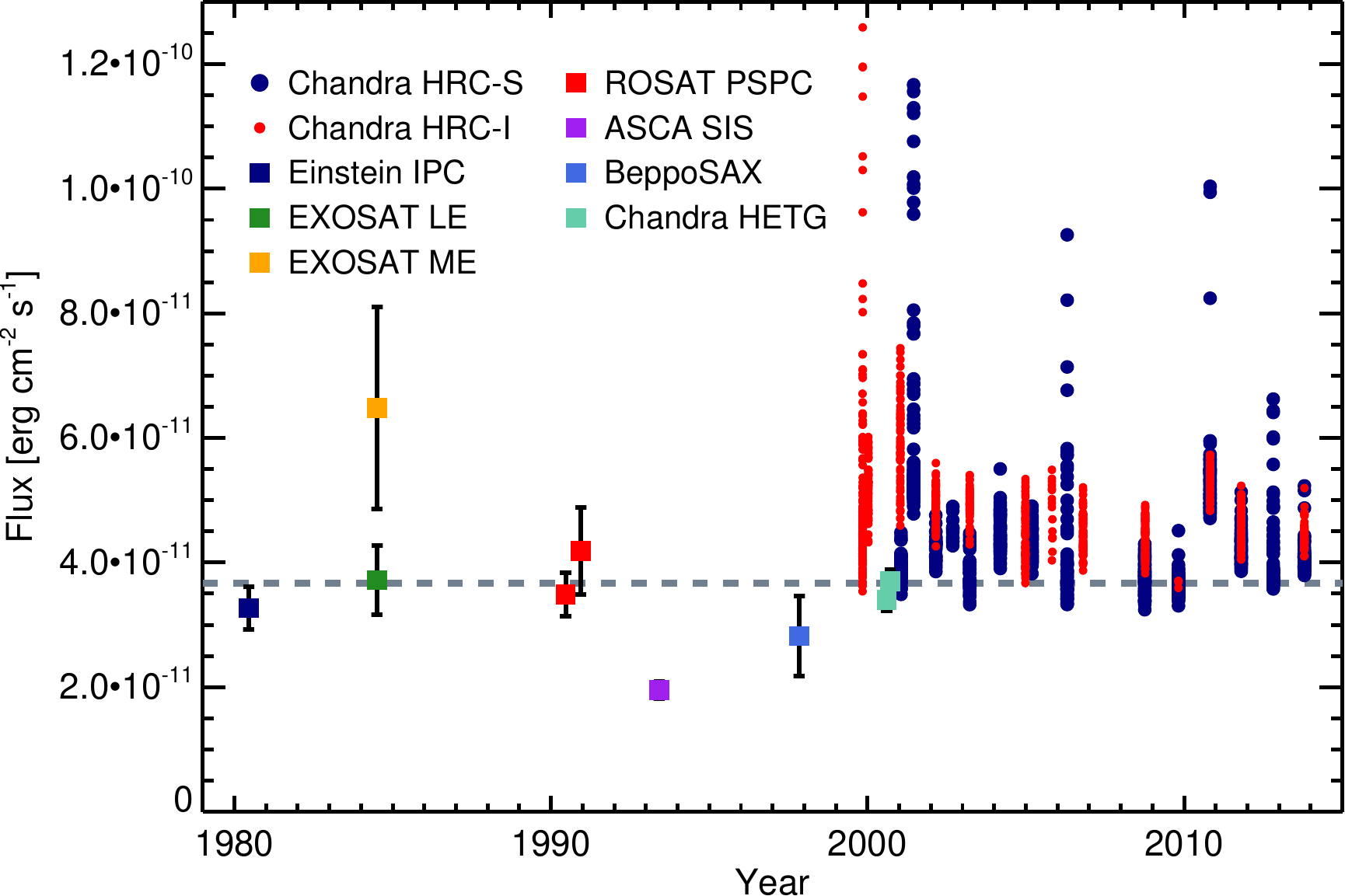}
\end{center}
\caption{
\chandra\ HRC-S and HRC-I X-ray light curves in the context of earlier X-ray observations of AR~Lac.  HRC data have been filtered to exclude primary eclipse within the phase range 0.8-1.2, but flares have been retained.  Earlier observations are summarized in Table~\ref{t:xrayvstime}.  The dashed horizontal line represents the mean of the observations prior to \chandra.  The mean quiescent X-ray flux from AR~Lac has remained consistently at $3.7\times 10^{-11}$~ergs~cm$^{-2}$~s$^{-1}$, to within about 10\%\ if the \asca\  SIS point is discarded, for the past 33 years.  
\label{f:xrayvstime}}
\end{figure*}

The HRC-I and HRC-S light curves as a function of time, in units of erg~cm$^{-2}$~s$^{-1}$, are illustrated in Figure~\ref{f:xrayvstime} 
in the context of the fluxes observed by previous X-ray missions, beginning with \einstein\ observations in 1980 June 14 about thirty three years ago.  The \chandra\ data were filtered to only include data in the phase range 0.2--0.8, outside of primary eclipse, but flares have been retained.  Flux levels and uncertainties for other missions are summarised in Table~\ref{t:xrayvstime}, and were based on count rates essentially determined by eye from figures in the relevant publications in which the data have appeared.  These count rates were converted to flux in the 0.5--5~keV band by folding an AR~Lac spectral model through the appropriate effective area curve obtained from the Portable Interactive Multi-Mission Simulator (PIMMS) database.\footnote{http://heasarc.nasa.gov/docs/software/tools/pimms.html}  The AR~Lac model was the same as that described in Section~\ref{s:chandra}.  We also performed sensitivity tests by changing the spectral model to the two-temperature best-fit model for the out of eclipse \asca\ observations analysed by \citet{Singh.etal:96}, and by halving the abundances of metals.  In all cases, the conversions from counts to flux changed by less than 10\% . 
 
The base flux level of AR~Lac---ie not considering flares---has been remarkably constant over the 13 years covered by {\it Chandra}, and varies by only 10\%\ or so.   Looking back further, over the 33 years of X-ray observations, the same base level is seen with only the {\it ASCA} observations from 1993 appearing significantly fainter than in the {\it Chandra} era by about 45\%.  The 1984 {\it EXOSAT} ME observation is only marginally consistent, but the instrument was only sensitive to X-ray energies above 1~keV and the derived flux is very sensitive to the adopted hot emission measure \citep[see, e.g.,][]{White.etal:90}.  The mean of the fluxes for the earlier missions, $3.7\times 10^{-11}$~erg~cm$^2$~s$^{-1}$, is in good agreement with the {\it Chandra} measurements,  while we also note that absolute calibration uncertainties of earlier missions could account for systematic differences at the 10\%\ or so or level. 

There is no evidence for any significant cyclic modulation, at least on the timescales covered by our observations.  The orbital period of AR~Lac has long been known to exhibit an oscillatory behaviour with a reported period of 35--50 years on top of a steady decline \citep{Hall.Kreiner:80,Van_Buren:86,Kim:91,Jetsu.etal:97, Lanza.etal:98,Qian.etal:99,Lu.etal:12}.  The origin of the oscillatory component remains uncertain though \citet{Lanza.etal:98} noted a possible relation with a $\sim 17$~yr surface spot cycle attributed to the K star and suggested the period variation could be due to the \citet{Applegate:92} mechanism that is driven by a magnetic cycle.  Since the G star appears to dominate the coronal emission, its cyclic behaviour would appear to be more relevant to this study than cycles on the K star.  The long baseline of the X-ray data presented here cannot rule out magnetic cycles, but any such cycle with a period of 17 or 35 years has very little influence on the X-ray coronal energy output.  

\citet{Kashyap.Drake:99} investigated the X-ray emission of active binary stars observed at various epochs by the \einstein\ and \rosat\ satellites and found that fluxes differed by 30--40\%\ or so on average between different epochs.  That study could only examine data averaged over whole observations, or in the case of \rosat, averaged over the all-sky survey, and so any flares would be implicitly included in the averages.  Figure~\ref{f:xrayvstime} demonstrates that typical base level emission variations are likely to be significantly smaller.  Relatively constant basal emission over decade timescales also appears to characterize the young K1 dwarf AB Dor \citet{Lalitha.Schmitt:13} despite evidence for an optical cycle, and the low-mass flare star VB8 \citep{Drake.etal:96}, and appears to be a general characteristic of very active stars.




\section{SUMMARY AND CONCLUSIONS}
\label{s:sum}

We have analysed an extensive set of {\it Chandra} HRC observations of the eclipsing RS~CVn-type binary AR~Lac obtained over a 13 year period and combined these data with observations by {\it EUVE}, {\it ASCA}, {\it ROSAT}, {\it EXOSAT} and {\it Einstein} that go back to 1980.  We find the quiescent base level coronal emission to be remarkably constant, with typical variations of 15\%\ or less over a period of 33 years.  Multi-orbit {\it Chandra} and {\it EUVE} observations indicate that stochastic variability likely dominates rotationally-modulated variability on orbital timescales.  Consequently, reconstructions of the spatial distribution of emitting plasma should be treated with caution.  Primary eclipses, when the more compact G2~IV component lies behind the K0 subgiant, are regularly detected but obvious secondary eclipses are absent.  Spherically-symmetric coronal models fitted to the {\it Chandra} and {\it EUVE} light curves cannot constrain the K star coronal scale height, but indicate a coronal scale height on the G component of $1.3 R_\odot$, or $0.86 R_\star$, and that the G star dominates the emission by a factor of 2--5.

\acknowledgments

JJD, VK, PR and BJW were funded by NASA contract NAS8-03060 to the {\it
Chandra X-ray Center} (CXC) and
thank the CXC director, H.~Tananbaum, and the CXC science team
for advice and support.  DPH was supported by Smithsonian Astrophysical Observatory contract SV3-73016 to MIT for Support of the Chandra X-Ray Center.


\end{document}